\documentclass[11pt]{article}
\usepackage{amsfonts, amsmath, amssymb}
\usepackage{array, booktabs}
\usepackage[margin=1in]{geometry}
\usepackage{multicol}
\usepackage[svgnames]{xcolor}  
\usepackage{setspace}
\usepackage{lineno}
\usepackage{siunitx} 
\usepackage{nameref}

\usepackage{graphicx}
\usepackage[margin=1cm]{caption}
\DeclareSIUnit\angstrom{\protect \text {Å}}

\usepackage[super]{cite}
\newcommand{\tz}{$[0 0 0 1]$}
\newcommand{\tx}{$[1 \bar{1} 0 0]$}
\newcommand{\ty}{$[1 \bar{2} 1 0]$}
\newcommand{\gba}{$\{ 3 1 \bar{4} 0 \}$}
\newcommand{\gbb}{$\{ 2 1 \bar{3} 0 \}$}
\newcommand{\gbc}{$\{ 1 1 \bar{2} 4 \}$}
\newcommand{\gbd}{$\{ 11\,2\,\overline{13}\,0 \}$}
\usepackage[hidelinks]{hyperref}

\begin{document}


\begin{center}
    {\large 
    \textbf{Grand canonically optimized grain boundary phases in hexagonal close-packed titanium}
    \vspace{0.2in}}
    
    {Enze Chen$^{1,2,3,4,*}$, Tae Wook Heo$^{2}$, Brandon C. Wood$^{2}$, Mark Asta$^{1,3}$, Timofey Frolov$^{2, *}$}
\end{center}

{\noindent \fontsize{10}{12}\selectfont 
$^1$ Department of Materials Science and Engineering, University of California, Berkeley, CA 94720, USA \\
$^2$ Materials Science Division, Lawrence Livermore National Laboratory, Livermore, CA 94550, USA \\
$^3$ Materials Sciences Division, Lawrence Berkeley National Laboratory, Berkeley, CA 94720, USA \\
$^4$ Present address: Department of Materials Science and Engineering, Stanford University, Stanford, CA 94305, USA \\
Email: 
$^{*}$enze@stanford.edu; 
$^{*}$frolov2@llnl.gov
}

\linenumbers 
\doublespacing
\begin{abstract}
\normalsize
Grain boundaries (GBs) profoundly influence the properties and performance of materials, emphasizing the importance of understanding the GB structure and phase behavior.
As recent computational studies have demonstrated the existence of multiple GB phases associated with varying the atomic density at the interface, we introduce a validated, open-source {GRand} canonical {Interface} {Predictor} (GRIP) tool that automates high-throughput, grand canonical optimization of GB structures.
While previous studies of GB phases have almost exclusively focused on cubic systems, we demonstrate the utility of GRIP in an application to hexagonal close-packed titanium. 
We perform a systematic high-throughput exploration of tilt GBs in titanium and discover previously unreported structures and phase transitions. 
In low-angle boundaries, we demonstrate a coupling between point defect absorption and the change in the GB dislocation network topology due to GB phase transformations, which has important implications for the accommodation of radiation-induced defects.

\end{abstract}

\newpage
\section*{Introduction} \label{sec:intro}

Grain boundaries (GBs) are interfacial defects in crystalline materials that have long been studied for their influence on materials properties and performance~\cite{sutton_1995}.
Given their ability to exist in multiple stable and metastable states, which have been termed GB phases~\cite{frolov_2015} or complexions~\cite{cantwell_2014}, it is desirable to obtain an atomic-level understanding of the GB structures and possible phase transition pathways between them~\cite{frolov_2013_gbphase, cantwell_2020}.
The structure--property relationships of these interfacial phases are believed to have a profound influence on an array of phenomena, such as diffusion~\cite{frolov_2013_diffuse} and GB migration~\cite{wei_2021} in materials.

Recent experiments have provided direct~\cite{meiners_2020} and indirect~\cite{rajeshwari_2020} evidence for GB phase stability, coexistence, and transitions in metallic systems;
however, given the vast five-dimensional space characterizing the macroscopic degrees of freedom (DOF) for GBs, it is not yet clear where these phases may appear.
Atomistic simulations provide a powerful tool to guide such searches and unveil the microscopic mechanisms underlying the formation of GB phases~\cite{frolov_2016}.
Previous atomistic modeling studies have discovered a diverse array of GB phases present in face-centered cubic (FCC)~\cite{frolov_2013_gbphase, zhu_2018, brink_2023}, body-centered cubic (BCC)~\cite{frolov_2018, frolov_2018_W}, diamond cubic~\cite{vonalfthan_2006, zhang_2009}, and other cubic systems~\cite{chua_2010, mazitov_2021}.
One notable feature shared by the aforementioned works is the ability to add or remove atoms from the GB region in the simulation cell, i.e., grand canonical optimization (GCO), which was required to access new ground states and metastable states.
While the exchange of atoms at an interface could naturally occur in real polycrystalline materials due to diffusion, irradiation, and mechanical deformation at finite temperature, this variation is omitted in the majority of computational simulations employing the $\gamma$-surface method~\cite{mishin_1998}.
The $\gamma$-surface method is the traditional technique for simulating GBs where only relative translations are allowed between two bulk slabs before the atoms are relaxed using conjugate gradient energy minimization to their equilibrium positions at \SI{0}{\kelvin}.
It is often adopted for its simplicity, but the deficiencies exposed by the previous studies suggest that more DOF must be considered during optimization in order to find the true ground-state structure in certain GBs.
A few alternative approaches from the literature for atomistic modeling of GBs include high-temperature molecular dynamics (MD) simulations~\cite{frolov_2013_gbphase, vonalfthan_2006, han_2016}, Monte Carlo sampling~\cite{banadaki_2018, yang_2018}, and evolutionary algorithms~\cite{chua_2010, restrepo_2013, yang_2018, zhu_2018, yang_2020} that can access a greater diversity of structures and atomic densities at the interface, with a concomitant trade-off in computational complexity.
These algorithms have employed GCO for GB structures in a variety of systems, although seldom in a high-throughput manner~\cite{banadaki_2018, han_2016}, and it remains unclear if the ubiquity of GB phases that they have yielded extends to lower-symmetry crystalline systems that are ubiquitous in nature and engineering applications.

A particular system of immense technological relevance is the hexagonal close-packed (HCP) crystal structure, which is considerably more complex than cubic systems, as it displays anisotropy in its crystalline lattice vectors and a basis containing more than one atom.  
This structure is adopted by elemental metals such as Mg, Zr, and Ti, the last of which ($\alpha$-Ti) will be the focus of this work.
Ti alloys are important structural alloys for aerospace, biomedical, and energy applications, particularly where high specific strength and strong corrosion resistance are desired~\cite{lutjering_2007}.
The importance of GBs in the $\alpha$-Ti system is highlighted in recent studies that used grain refinement to mitigate low-temperature oxygen embrittlement in $\alpha$-Ti~\cite{chong_2023} and 3D electron backscatter diffraction to map out the complete GB character distribution in this system~\cite{kelly_2016}.
In a previous study~\cite{hooshmand_2021}, we used an evolutionary algorithm to discover a ground-state structure for the \gbc\tx\ twin boundary (TB) in $\alpha$-Ti that was in closer agreement to density functional theory (DFT) calculations and high-resolution transmission electron microscopy results than previously reported structures.
In addition to these experimental works, there are also several atomistic simulation studies in the literature that systematically model symmetric tilt grain boundaries (STGBs) in $\alpha$-Ti along the \tz~\cite{wang_1997, zheng_2017}, \tx~\cite{farkas_1994, wang_2012b, ni_2015, bhatia_2013}, and \ty~\cite{wang_2012a, bhatia_2013} tilt axes.
Despite the simplicity of STGBs---only a tilt axis and tilt angle ($2 \theta$) are required to describe the crystallographic misorientation between two bulk crystals---they include coherent TBs as an important subclass, several of which are experimentally observed in deformation microstructures and thus important for mechanical behavior in $\alpha$-Ti~\cite{lutjering_2007, chong_2023, hooshmand_2021}.
STGBs are also model systems to study the geometric relationships of defects at the interface~\cite{wang_2012a};
however, as the previous studies utilized the $\gamma$-surface method, it is important to clarify the effects of GCO on STGB structure in $\alpha$-Ti and more broadly whether interfacial phases exist in HCP metals.

Herein, we perform GCO of low-index STGBs in $\alpha$-Ti using an open-source {GRand} canonical Interface Predictor (GRIP) tool that we developed to rigorously sample microscopic DOF at the GB.
We use this tool along with empirical potentials to perform GB structure search, discovering new ground-state structures and GB phases.
We further employ high-temperature MD simulations to explore the \gbb\tz\ STGB and demonstrate GB phase (meta)stability and phase transitions through a novel dislocation-pairing mechanism.
We conclude by discussing the broader implications of these results on GB phase behavior in HCP systems and how the GRIP tool can benefit future studies for diverse crystal structures.

\section*{Results}  \label{sec:results} 

\subsection*{Grand canonical optimization---the GRIP tool}

GB structure prediction is a long-standing challenge in materials modeling that requires rigorous and often advanced sampling of possible interfacial structures. 
Previous studies of GBs in HCP metals generated the interfaces using the common $\gamma$-surface method~\cite{mishin_1998}, which is not guaranteed to yield the true ground state configuration in general~\cite{vonalfthan_2006, zhu_2018}.
In the traditional approach, conjugate gradient minimization  from different starting points representing distinct relative transitions of the grains across the boundary simply allows the atoms to fall into a nearby local minimum, which may be far away from the ground state. 
For example, complex GB core configurations may exist that require significant rearrangement of the constituent atoms~\cite{vonalfthan_2006, zhu_2018}.

The other significant limitation of the $\gamma$-surface method is that it is not grand canonical: All GBs created using this method are composed of the same number of atoms derived from the constituent perfect half-crystals. 
This poses a substantial constraint because many other structures, including true ground states, can be realized out of a different number of atoms at the interface~\cite{vonalfthan_2006, frolov_2013_gbphase, frolov_2018}.
For STGBs and a fixed reconstruction area, the number of distinct atomic densities that can give rise to different GB structures is given by the total number of atoms in one atomic plane parallel to the boundary, which we denote $N^{\mathrm{bulk}}_{\mathrm{plane}}$.
This quantity is the limit because removing a full plane of atoms from a crystal will return the exact same configuration up to a relative grain translation.

\begin{figure}[!ht]
    \centering
    \includegraphics[width=0.7\linewidth]{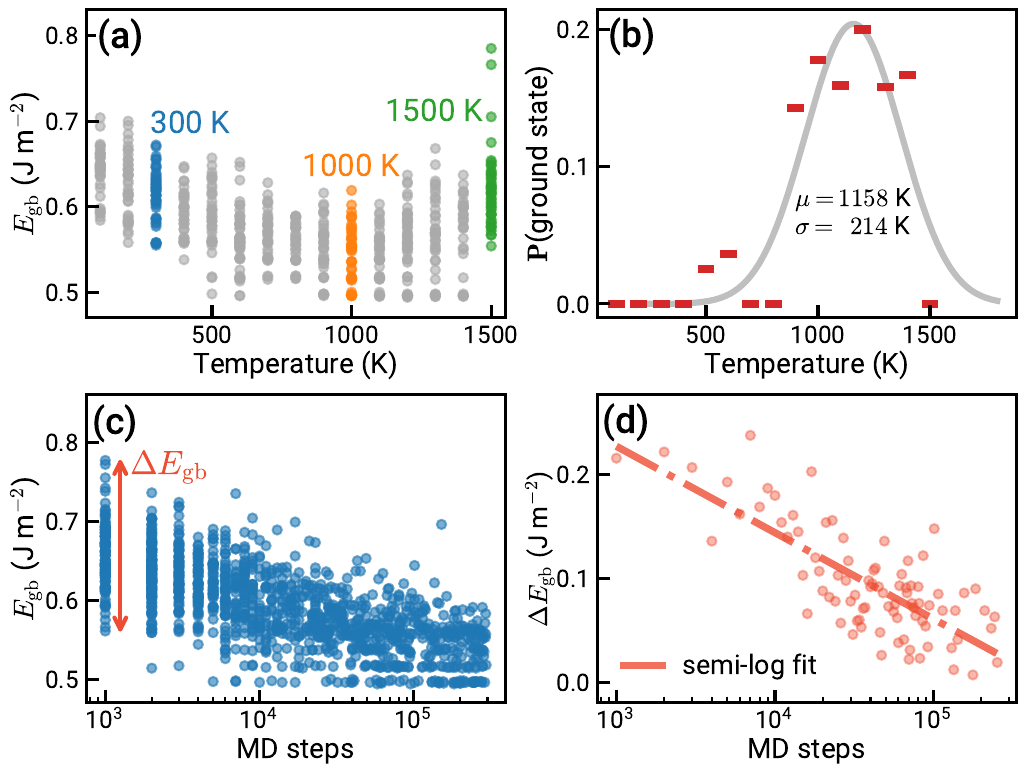}
    \caption{\textbf{Success rate as a function of search parameters in GRIP}.
    The success rate of finding the ground state is sensitive to the search parameters, including temperature and duration of the MD sampling. 
    The optimal parameters are not known a priori and vary for each particular boundary and reconstruction, demonstrating why rigorous sampling of the parameter space is critical.
    (a) The U shape of the $E_{\mathrm{gb}}$ vs. $T$ plot illustrates the inefficient frozen dynamics of the GB structure at low temperatures and the generation of disordered liquid-like GBs at very high temperatures for a representative GB. 
    (b) The fraction of ground-state structures out of all structures sampled at each temperature are plotted and fitted with a Gaussian distribution.
    (c) A sufficiently large number of MD steps is required to obtain the ground state, even at an optimal $T$. 
    (d) $\Delta E_{\mathrm{gb}}$ is plotted against the number of MD steps.
    The least-squares regression line shows the convergence in energy at longer duration.
    }
    \label{fig:scan_MD}
\end{figure}

Here we address these shortcomings through the development of an open-source tool GRIP to perform grand canonical GB structure search.
During the optimization, we systematically explore all possible microscopic DOF by sampling different relative grain translations and atomic densities (see \nameref{sec:methods} for details). 
The latter is accomplished by randomly removing a fraction of atoms between $0$ and $N^{\mathrm{bulk}}_{\mathrm{plane}}$ from the boundary plane. 
For a fixed translation and number of GB atoms, we optimize the GB structure using dynamic sampling (performed here using MD simulations) at different temperatures within a wide window between room temperature and \SI{1200}{\kelvin} (approximately $T_{\alpha \rightarrow \beta}$ for Ti), and for different durations up to \SI{0.6}{\nano\second}. 
At the end of each MD run, we perform conjugate gradient energy minimization at \SI{0}{\kelvin} until convergence before calculating the GB energy, $E_{\mathrm{gb}}$ (see \autoref{eq:Egb} in \nameref{sec:methods}). 
The random and diverse GB structure initialization coupled with the extensive dynamic sampling for each microscopic DOF done by hundreds of parallel calculations ensure a rigorous GB structure exploration.

To further underscore the need for rigorous sampling, \autoref{fig:scan_MD} shows the structural diversity and success rates from randomly sampling MD simulation parameters, namely temperature ($T$) and duration (MD steps).
\autoref{fig:scan_MD}a shows the range of $E_{\mathrm{gb}}$ as a function of $T$ when the duration is fixed, and only at intermediate temperatures does the algorithm find the ground-state structure. For this representative boundary, the U shape of the $E_{\mathrm{gb}}$ vs. $T$ plot illustrates the inefficient frozen dynamics of the GB structure at low temperatures and the generation of disordered liquid-like GBs at very high temperatures.
From this data, we can compute the probability of finding the ground state---calculated as the fraction of ground-state structures out of all sampled structures at each $T$---which peaks at approximately \SI{1150}{\kelvin} and is zero for very low and very high temperatures (\autoref{fig:scan_MD}b).
These panels illustrate the existence of an optimal $T$ that is sensitive to the structural DOF of each system and outside of which the GB may fail to be optimized.
Analogously, simply choosing an optimal $T$ (e.g., \SI{1000}{\kelvin}) is insufficient, as too few MD steps will never achieve the ground state, as shown in \autoref{fig:scan_MD}c.
The energy range, $\Delta E_{\mathrm{gb}}$, is plotted in \autoref{fig:scan_MD}d to show how the spread generally decreases as the MD duration increases;
however, we emphasize that the optimal parameters are not known a priori. These optimal parameters can vary significantly not only with the boundary character described by the five macroscopic DOF, but also for larger area reconstructions of the same boundary. The uniform sampling of possible optimization parameters and GB DOFs implemented in GRIP allows for robust high-throughput optimization of large GB datasets.

As motivated in the Introduction, we showcase the performance of GRIP in the following sections through a detailed analysis of STGBs in HCP $\alpha$-Ti.
Importantly, however, we note that we also comprehensively benchmark our tool by reproducing well-studied literature results for tilt and twist GBs in elemental cubic metals~\cite{frolov_2013_gbphase, frolov_2018_W} and more challenging covalently-bonded, lower-symmetry systems~\cite{vonalfthan_2006, banadaki_2018} 
(Supplementary \autoref{fig:gcoval1}).
Even in the thoroughly studied BCC W system~\cite{frolov_2018_W}, we discover a new ground-state structure with a different GB atomic density and markedly different dislocation network in the GB 
(Supplementary \autoref{fig:gcoval2}).
Such results, while not discussed further in this work, underscore the opportunities of having a robust method for exploring GB phase space across disparate chemical systems.
The new ground states also position GRIP as a tool capable of advancing the state of the art in GB structure prediction through its extensive dynamic sampling of the relevant DOF.

\subsection*{Survey of new GB phases in HCP $\alpha$-Ti}

\begin{figure}[!ht]
    \centering
    \includegraphics[width=0.5\linewidth]{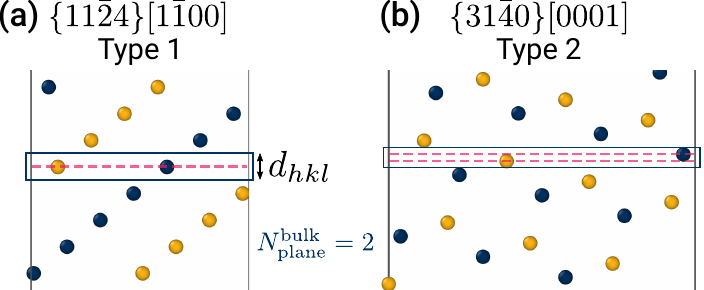}
    \caption{\textbf{Calculation of the number of atoms per plane, $N^{\mathrm{bulk}}_{\mathrm{plane}}$, in HCP}. 
    The accurate calculation of $N^{\mathrm{bulk}}_{\mathrm{plane}}$ ensures that GB structures with all possible atomic densities are explored. 
    Because HCP crystals have two basis atoms, two different cases are possible when
    (a) all atoms inside the plane have the same $z$-position, or 
    (b) they have two distinct $z$-positions resulting in two structurally different surface terminations.
    In both cases, $N^{\mathrm{bulk}}_{\mathrm{plane}}$ is calculated as the total number of atoms found inside the region spanned by the hexagonal interplanar spacing, $d_{hkl}$ (boxed). 
    The distinct $z$-positions of atoms belonging to the same plane are indicated by the dashed magenta lines.
    }
    \label{fig:terminations_Nplane}
\end{figure}

The two-atom basis of HCP Ti presents additional considerations during optimization, and \autoref{fig:terminations_Nplane} illustrates one nuance in having two possible cases of calculating $N^{\mathrm{bulk}}_{\mathrm{plane}}$. 
For the orientation shown in \autoref{fig:terminations_Nplane}a, all atoms found inside the planar region have the same $z$-positions indicated by the dashed magenta line. 
Such orientations are analogous to cubic systems and have only one distinct surface termination.
For the second case shown in \autoref{fig:terminations_Nplane}b, the atoms belonging to the same plane can have two distinct $z$-positions, giving rise to two structurally different surface terminations.
These two distinct terminations are possible because HCP has two basis atoms.
In all orientations, the thickness of the planar region, $d_{hkl}$, corresponds to the smallest normal component of a lattice vector connecting two atoms on the same sublattice with different $z$-positions.
We further note that this definition of a plane of atoms works for both cases, allowing us to uniformly apply it in calculating GB atomic density, $n$ (see \autoref{eq:n} in \nameref{sec:methods}).

\begin{figure}[!ht]
    \centering
    \includegraphics[width=\linewidth]{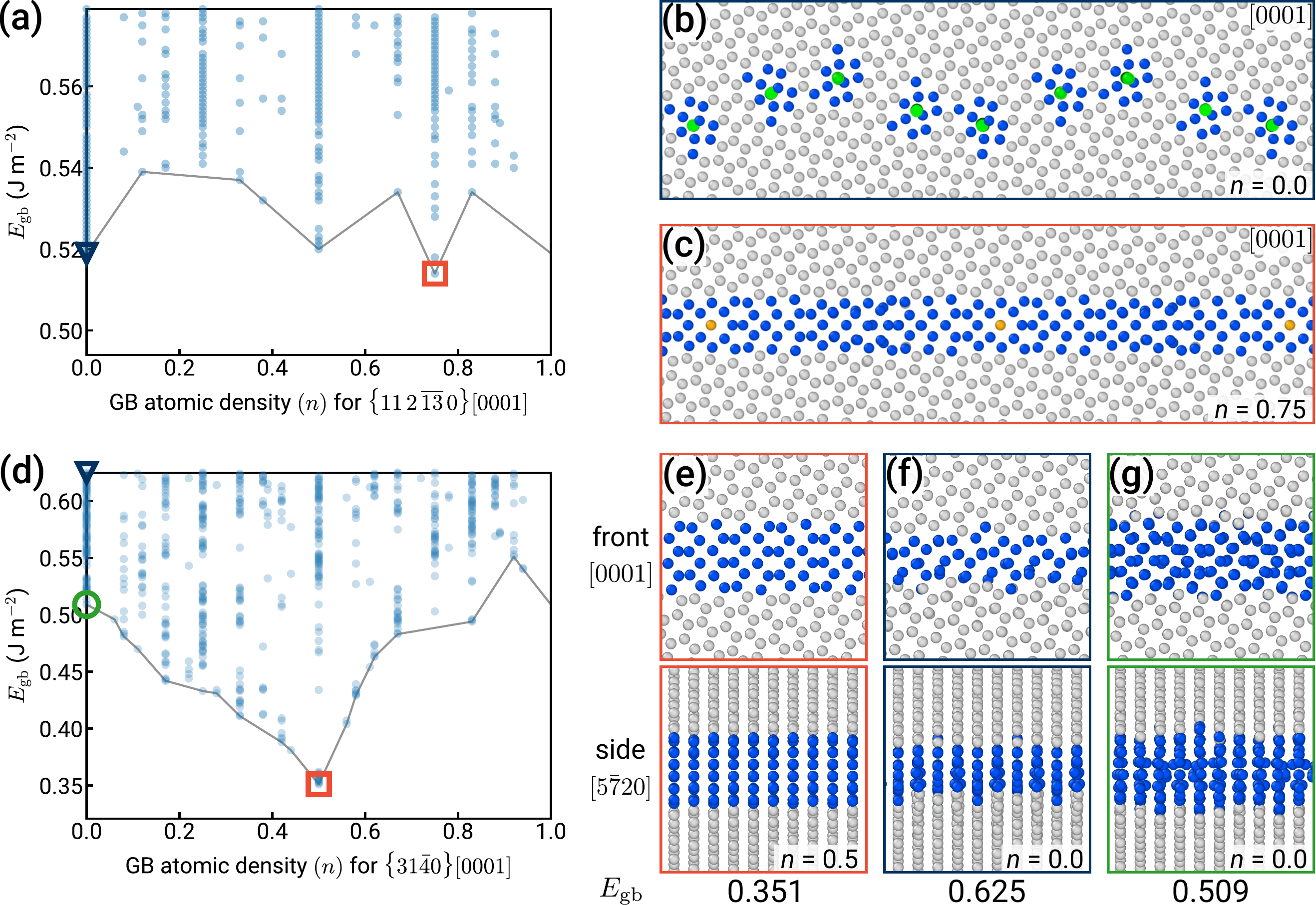}
    \caption{\textbf{Representative GB structure searches using GRIP}.
    (a) GB energy $E_{\mathrm{gb}}$ vs. GB atomic density $n$ reveals two GB phases of $\Sigma 49$\gbd\tz\ which are shown in panels (b) and (c). 
    (b) The $n=0$ GB phase, composed of edge dislocations, is metastable at 0 K. 
    (c) The $n = 0.75$ GB phase is the ground state, highlighting the importance of GCO.
    (d) $E_{\mathrm{gb}}$ vs. $n$ for $\Sigma 13$\gba\tz.
    On the right, two orthogonal projections are shown for each minimum-energy structure obtained using 
    (e) GRIP at $n=0.5$,
    (f) the $\gamma$-surface method, and
    (g) GRIP at $n=0$.
    The atoms are colored according to the common neighbor analysis (CNA) in OVITO~\cite{stukowski_2012_cna, stukowski_2009}.
    }
    \label{fig:gco_plot}
\end{figure}

\autoref{fig:gco_plot} shows the results of the GRIP searches for two representative boundaries evaluated using a modified embedded-atom method (MEAM) potential~\cite{hennig_2008}, illustrating the need for grand canonical structure optimization for GBs in HCP Ti. 
Panels (a) and (e) show plots of $E_{\mathrm{gb}}$ vs. $n$, which was introduced for cubic crystals in our previous work~\cite{frolov_2013_gbphase, zhu_2018}. 
Each point on the plot corresponds to a particular GB structure obtained after energy minimization. 
The thorough exploration enabled by GRIP generates hundreds of distinct structures covering different densities and energies.

For the \gbd\tz\ GB, the structure search identifies two GB phases with different atomic densities $n=0$ and $n=0.75$.
The structures are illustrated in panels (b) and (c),  respectively. 
The $n=0$ phase does not require insertion or removal of atoms and is metastable at \SI{0}{\kelvin}. 
It is composed of well-separated cores of edge dislocations with Burgers vectors $\mathbf{b}_{\text{I}} = \frac{1}{3} \langle 1 \bar{2} 1 0 \rangle$, as identified in green by the dislocation extraction algorithm (DXA) in OVITO~\cite{stukowski_2012_dxa, stukowski_2009}. 
The newly predicted ground state of this boundary has $n=0.75$ and thus cannot be generated by using the simplistic $\gamma$-surface approach or sampling different terminations. 
Its structure is significantly different from the $n=0$ state, where the dislocation cores overlap and the boundary structure appears completely flat. 
The energy of the ground state (corresponding to $n=0.75$) is \SI{1}{\percent} lower than that of the metastable phase ($n=0$).
We finally note that this GB is a single-surface termination type of boundary where all atoms belonging to a bulk plane have the same $z$-coordinate.

\begin{figure}[!b]
    \centering
    \includegraphics[width=0.5\linewidth]{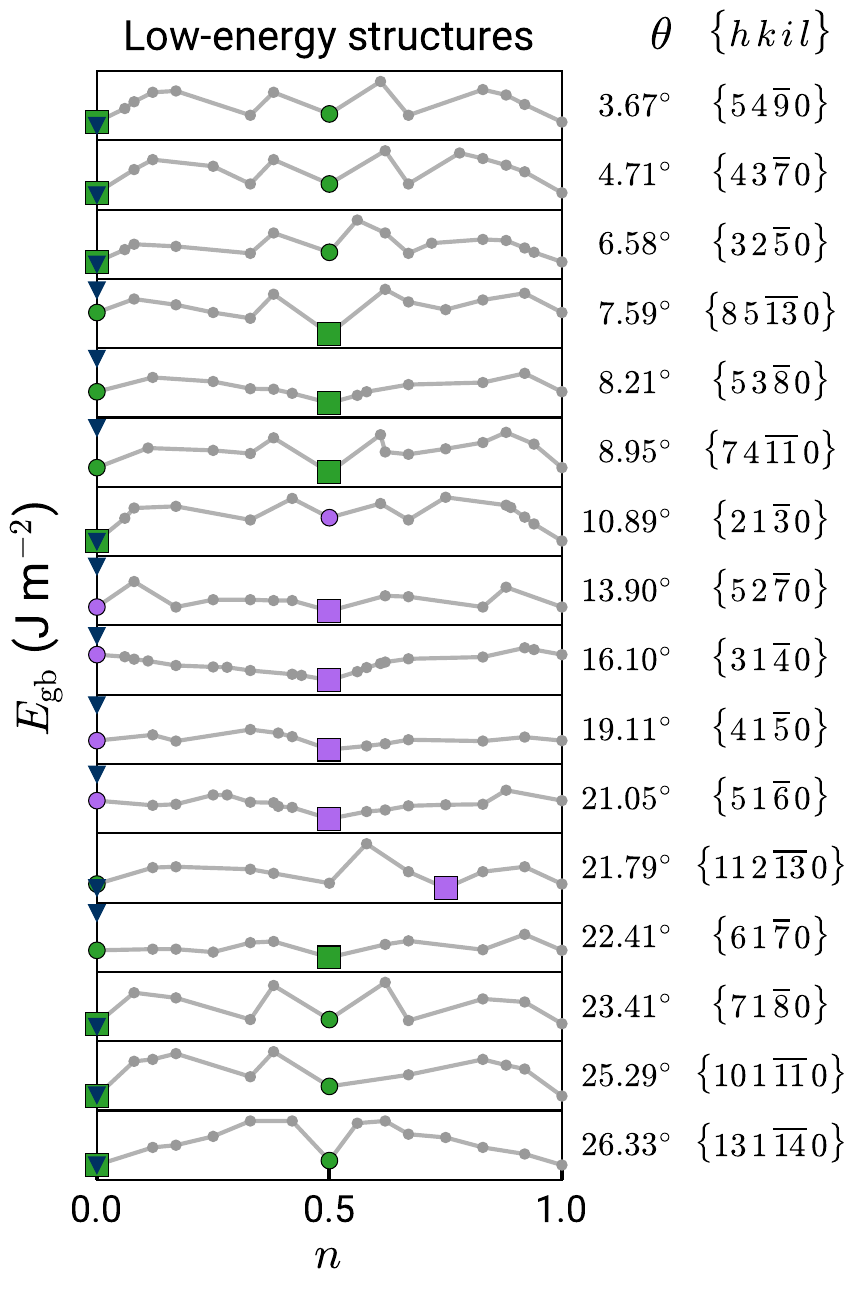}
    \caption{\textbf{GB energy map of \tz\ STGBs}.
    GRIP finds new  ground states and multiple GB phases within the entire misorientation range, demonstrating the need for GCO.
    Each subplot is analogous to the boundary in \autoref{fig:gco_plot}a, marking the minimum-energy structures at different GB atomic densities.
    The squares mark the ground state for each tilt angle, the larger circles mark a metastable state, and the blue triangles mark the $\gamma$-surface structure.
    The green and purple marker colors correspond to different GB phases (commensurate with the dislocation core colors in \autoref{fig:phases}).
    }
    \label{fig:Eminima}
\end{figure}

The GRIP search for the \gba\tz\ GB shown in \autoref{fig:gco_plot}d illustrates that the microscopic descriptor $n$ properly captures all possible distinct GB configurations, even for orientations with two distinct surface terminations. 
Similar to the previous boundary, the prediction of the ground-state structure at $n=0.5$ also requires an insertion (or removal) of half of the atoms in one \gba\ plane;
however, different from the first example, this particular ground-state structure can also be generated using the $\gamma$-surface approach that considers two possible surface terminations.
The different surface terminations are obtained in a straightforward manner by removing half of a plane that contains two layers of atoms, as visualized in 
Supplementary \autoref{fig:terminations}.
We emphasize that while sampling terminations may suffice in some cases, it is clearly restricted to atomic densities of $0$ and $0.5$, thereby performing very limited optimization of the atomic structure.
In our search, for example, the GRIP tool finds a GB structure at $n = 0$ with $E_{\mathrm{gb}} = \SI{0.509}{\joule\per\meter\squared}$ (green circle), approximately \SI{18}{\percent} lower in energy than the best $\gamma$-surface structure (blue triangle).
For comparison, the ground-state structure and the two metastable states at $n=0$ are shown in panels (e), (f), and (g).

\begin{figure}[!ht]
    \centering
    \includegraphics[width=0.5\linewidth]{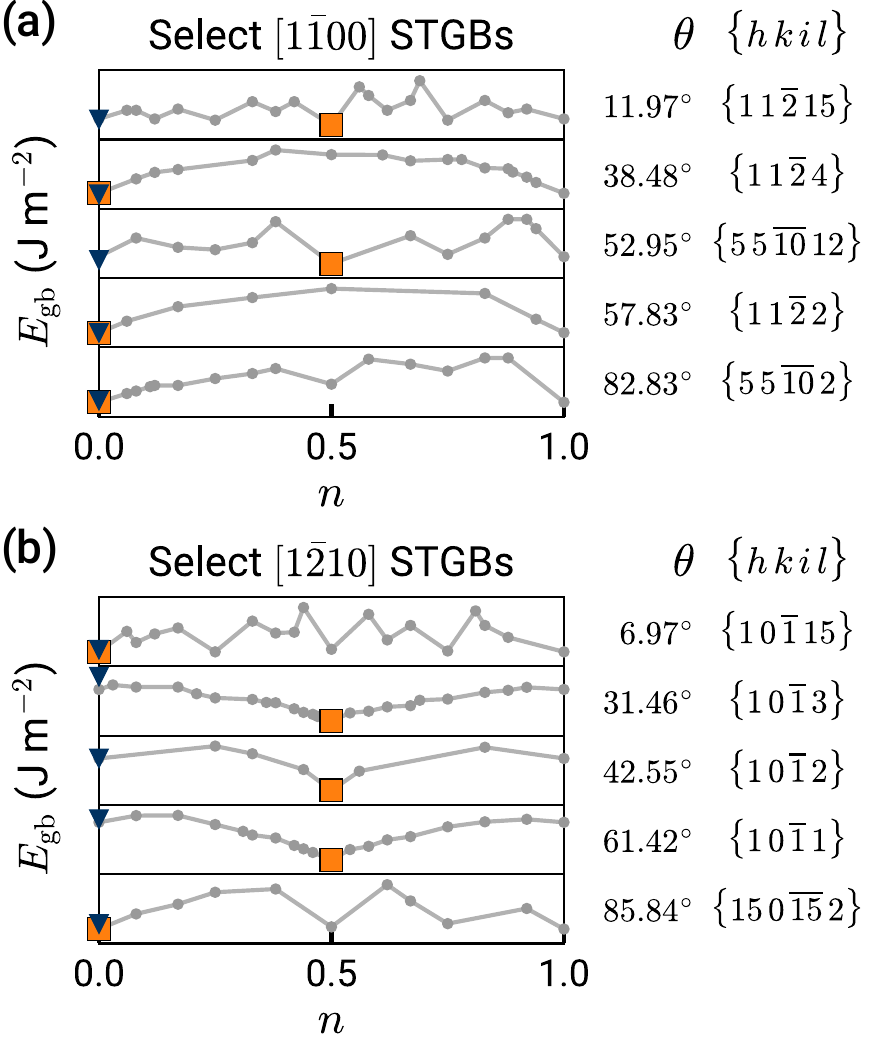}
    \caption{\textbf{GB energy maps of (a) \tx\ and (b) \ty\ STGBs}.
    Five different misorientations are selected for each tilt axis as representative boundaries. 
    Minima at $n=0.5$ indicate that GCO is broadly required to find the ground states in all families of STGBs studied in $\alpha$-Ti.
    }
    \label{fig:Eminima_select}
\end{figure}

Our structure searches performed for 150 GBs with three different tilt axes show that the need for GCO and presence of multiple GB phases is a general phenomenon in HCP Ti. 
\autoref{fig:Eminima} summarizes the results from GRIP for the family of \tz\ STGBs studied. 
Each subplot is equivalent to the gray boundary in \autoref{fig:gco_plot}a, denoting the minimum-energy structures at different $n$ and the square marks the ground state. 
Evidently as many of the minima are located at $n = 0.5$, GCO is necessary to find the ground state in multiple \tz\ STGBs.
Similar to \gba, the $\gamma$-surface method often performs poorly for these GBs, getting higher energies and different structures than the GRIP tool. 
The color map of the ground states reveals three distinct intervals that correspond to different GB structural units.
The low-angle GBs in the intervals $\theta \le \SI{6.58}{\degree}$ and $\theta \ge \SI{23.41}{\degree}$ are composed of isolated $\mathbf{b}_{\text{I}}$ edge dislocations (green markers) that for the lowest angles do not require GCO. 
The near-energy-degenerate minima at $n=0.5$ are composed of the same type of dislocations, with the extra atoms accommodated by dislocation climb, resulting in GB structures with unevenly spaced GB dislocations. 
Different GB dislocations stabilize at $\theta \approx \SI{10.89}{\degree}$ with twice the Burgers vector of $\mathbf{b}_{\text{II}} = \frac{1}{3} \langle 2 \bar{4} 2 0 \rangle$ (purple markers). 
We investigate the transition between these two states in this GB in detail in the next section. 
For high-angle GBs in the interval $\SI{13.90}{\degree} \le \theta \le \SI{21.79}{\degree}$, the ground states at $n=0.5$ are composed of structural units that match the dislocation core structures of $\mathbf{b}_{\text{II}}$, as outlined in 
Supplementary \autoref{fig:comp_units}. 
Additional energy maps for select \tx\ and \ty\ STGBs with low and high tilt angles out of 134 total studied are shown in \autoref{fig:Eminima_select}, and results for the embedded-atom method (EAM) potential~\cite{zope_2003} are presented in 
Supplementary Figures~\ref{fig:Eminima_Zope_1} and \ref{fig:comp_structs}, and \nameref{sec:si_comp}.
We also perform select DFT calculations using the optimized GRIP structures as inputs to confirm the stability of the GB dislocation core structures and the relative energies between phases (see 
Supplementary \autoref{fig:dft_val}).
Taken together, these results demonstrate the ubiquitous need for grand canonical sampling in locating the ground-state structures for multiple tilt axes in HCP Ti.

\subsection*{Phase transitions and coexistence}

The multiple GB phases predicted by GRIP opens up an opportunity to explore GB phase transformations in HCP Ti;
specifically, we focus on low-angle STGBs and investigate transformations that change the topology of the dislocation network arrangement.
By elucidating the transformation pathways, we are able to predict the structure of a nucleus with a distinct dislocation network topology embedded inside a different parent dislocation network.
We use point defects to drive the transformation and we study the coupling between defect absorption and changes in the dislocation network topology.
While low-angle GB phase transformations due to solutes and temperature have been previously reported by experimental observations and simulations in a few metals~\cite{meiners_2020, sickafus_1987, frolov_2018_W, peter_2018}, the questions of transition states and the role of intrinsic point defects have not been investigated.

\begin{figure}[!ht]
    \centering
    \includegraphics[width=\linewidth]{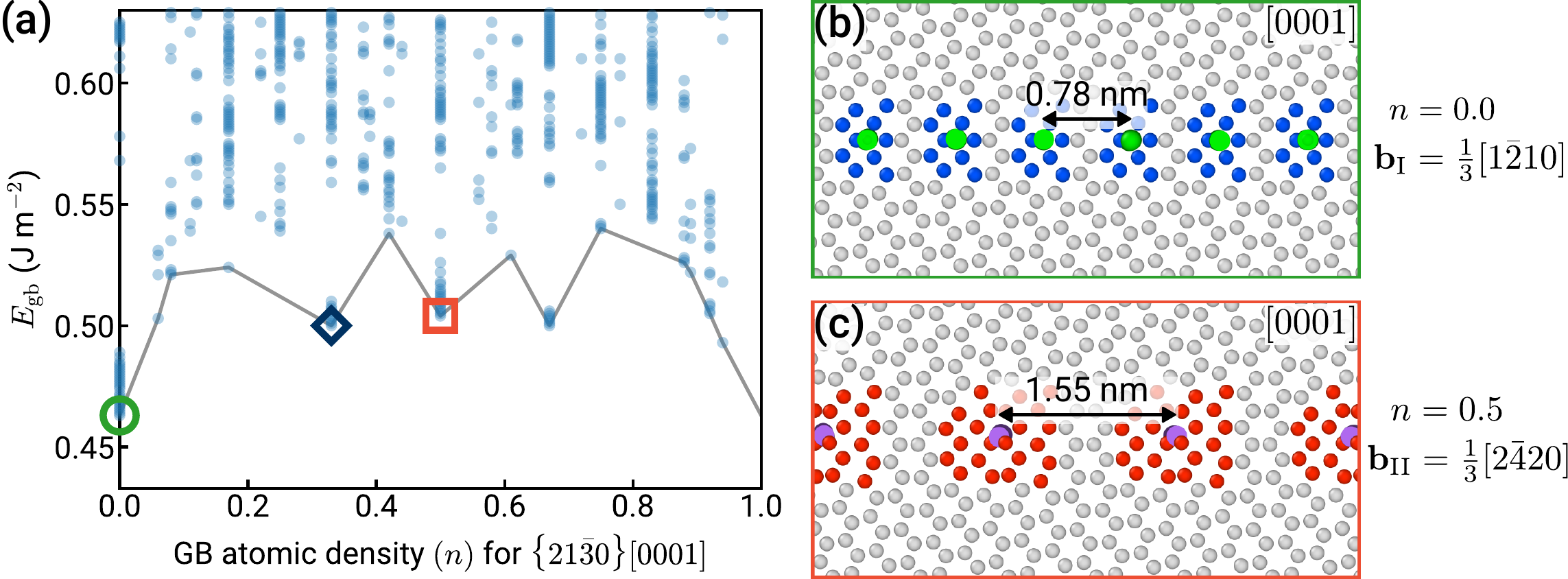}
    \caption{\textbf{GB phases of $\Sigma 7$\gbb\tz\ STGB}.
    (a) The plot of $E_{\mathrm{gb}}$ vs. $n$ reveals two GB phases, with the ground state at $n=0$ and a metastable phase at $n=0.5$, shown in (b) and (c), respectively. 
    Both states are composed of edge dislocations indicated by green and purple lines (identified using DXA in OVITO~\cite{stukowski_2012_dxa, stukowski_2009}). 
    The Burgers vectors of the metastable boundary is twice that of the ground state. 
    The non-HCP atoms of the dislocation cores are colored according to the CNA.
    }
    \label{fig:phases}
\end{figure}

We select the \gbb\tz\ STGB, which marks the transition between the two different GB dislocation types at a misorientation angle of $\theta \approx \SI{10.9}{\degree}$.
The structure search performed on this GB is illustrated in \autoref{fig:phases}, where we identify two distinct GB phases corresponding to atomic densities $n=0$ (green circle) and $n=0.5$ (orange square). 
Both structures correspond to GB energy cusps with respect to $n$ and the structures of the two phases at \SI{0}{\kelvin} are shown in \autoref{fig:phases}b and \ref{fig:phases}c, respectively. 
The ground state ($n=0$) is composed of an array of $\mathbf{b}_{\text{I}} = \frac{1}{3} \langle 1 \bar{2} 1 0 \rangle$ edge dislocations, while the second phase is composed of $\mathbf{b}_{\text{II}} = \frac{1}{3} \langle 2 \bar{4} 2 0 \rangle$ edge dislocations with Burgers vector twice that of the ground state and consequently half the line density within the GB plane.
The optimized dislocation core structures are consistent with the `T' and `A' structural units, respectively, reported by Wang and Ye using constrained molecular statics~\cite{wang_1997}.
We perform MD simulations of each structure at temperatures as high as \SI{1150}{\kelvin} for up to \SI{20}{\nano\second} to confirm that they are dynamically stable and indeed represent two GB phases.
The other two energy cusps at $n=0.33$ and $n=0.67$ are the mixed states expected from the lever rule, where the GB region is patterned by weighted fractions of $\mathbf{b}_{\text{I}}$ and $\mathbf{b}_{\text{II}}$ dislocations corresponding to the proportions between $n=0$ and $n=0.5$, as shown in 
Supplementary \autoref{fig:phases_mixed}.

\begin{figure}[!ht]
    \centering
    \includegraphics[width=\linewidth]{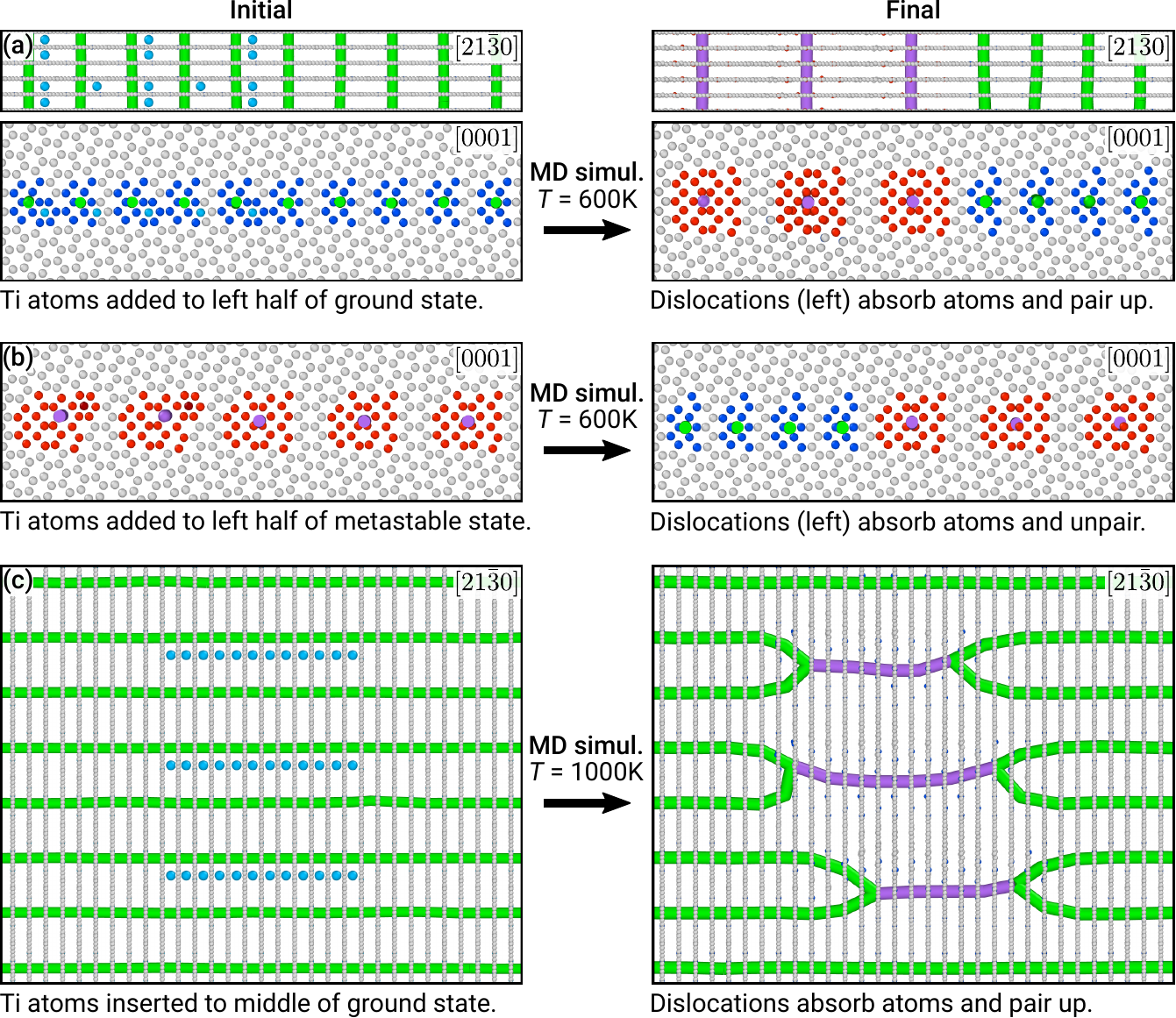}
    \caption{\textbf{Topological GB dislocation network transformation in \gbb\tz}.
    (a) Adding Ti atoms (light blue) to the left half of the ground-state structure and performing MD simulations at $T = \SI{600}{\kelvin}$ triggers a dislocation-pairing transformation $2 \mathbf{b}_{\text{I}} \rightarrow \mathbf{b}_{\text{II}}$ in a quasi-2D geometry.
    (b) Analogously, adding Ti atoms (dark red) to the left half of the metastable structure triggers a dislocation-unpairing transition $\mathbf{b}_{\text{II}} \rightarrow 2 \mathbf{b}_{\text{I}}$.
    (c) Topological transition of the GB dislocation network upon defect absorption. The view of the GB plane shows a paired-dislocation GB island (nucleus) inside the parent ground state.
    }
    \label{fig:phase_trans}
\end{figure}

Because the two GB phases are composed of different numbers of atoms, first-order transitions between the two structures can be triggered by changing the concentration of point defects~\cite{frolov_2013_gbphase}. 
The requisite high, local non-equilibrium concentrations of vacancies or interstitials may occur, for example, as a result of radiation damage, rapid cooling from high temperatures, or deformation by creep. 
To mimic these conditions, we insert extra atoms into interstitial sites in the ground-state structure, triggering a local transformation of the GB structure illustrated in \autoref{fig:phase_trans}. 
During the transformation, the $\mathbf{b}_{\text{I}}$ dislocations of the ground-state structure pair up into $\mathbf{b}_{\text{II}}$ dislocations and absorb the extra atoms. 
Analogously, adding Ti atoms to the left half of the metastable structure and performing high-temperature MD triggers a dislocation-unpairing transition ($\mathbf{b}_{\text{II}} \rightarrow 2 \mathbf{b}_{\text{I}}$) as shown in \autoref{fig:phase_trans}b. 
The transformed states remain stable at finite temperature and the transformation can be reversed by introducing vacancies near the GB, which we show in 
Supplementary \autoref{fig:phases_si}. 
Effectively, this sequence of states and partial transformations illustrate the possibility of GB transformation-mediated creep. 
Indeed, such a bicrystal can grow (shrink) by periodically alternating its GB structure and absorbing only half a plane of atoms (vacancies) at a time. 
If only one GB phase were present, the whole plane of atoms would have to be absorbed in concert through disconnection motion before returning to the original GB structure.

The simulated heterogeneous states containing two different GB phases show stable coexistence in the closed system at high temperatures. 
While not visible in \autoref{fig:phase_trans}a and \ref{fig:phase_trans}b, the two phases are separated by a line defect called a GB phase junction, which is a dislocation as well as a force monopole~\cite{winter_2022}. 
The Burgers vector of this junction is non-zero because the GB phases have different dimensions~\cite{frolov_2021}. 
The structure of this defect becomes more apparent when considering nucleation in fully 3D.
To illustrate the shape of the nucleus during such a transformation, we increase the cross-section of the GB and place interstitial atoms of Ti (light blue) in a relatively small section. 
During the subsequent high-temperature simulation at $T = \SI{1000}{\kelvin}$, the extra atoms diffuse to the boundary core and locally trigger the pairing transition. 
The equilibrium structure of the obtained nucleus is illustrated in \autoref{fig:phase_trans}c. 
The transformation changes the dislocation network topology as the dislocations of the parent structure shown in green ($\mathbf{b}_{\text{I}}$) pair up on the nucleus boundary to form three individual purple segments ($\mathbf{b}_{\text{II}}$).
GB phase nucleation by absorption of point defects has been previously investigated in high-angle boundaries~\cite{frolov_2013_gbphase, frolov_2018_W}.
The important distinction of the transformation studied here is that it occurs in a low-angle GB;
therefore, the core structure of the GB phase junction is represented by a collection of dislocation nodes where two dislocations pair up into one. 
To the best of our knowledge, such transition states facilitating the change in the GB dislocation network topology in a pure metal by point defect absorption have not been previously reported.

\section*{Discussion}  \label{sec:discuss}  

In this work, we perform grand canonical GB structure search to discover new GB phases in an HCP metal, $\alpha$-Ti.
While GBs in $\alpha$-Ti have been investigated extensively~\cite{wang_1997, zheng_2017, farkas_1994, wang_2012a, wang_2012b, ni_2015, bhatia_2013},
prior simulations were restricted to a fixed number of atoms derived from perfect surface terminations with no point defects (see \nameref{sec:si_comp}). 
By rigorously exploring atomic densities at GBs, we show that the minimum-energy structures can be found for atomic densities inaccessible to the $\gamma$-surface method for both high-angle and low-angle GBs across the misorientation range. 
The ubiquitous need for GCO and presence of multiple GB phases with different atomic densities is consistent with phenomena previously illustrated in elemental cubic metals with FCC~\cite{frolov_2013_gbphase, zhu_2018, brink_2023} and BCC~\cite{frolov_2018, frolov_2018_W} crystal structures.

Subsequent high-temperature MD simulations guided by this detailed sampling of phase space yield the discovery of a novel GB phase transformation mechanism.
The two phases shown in \autoref{fig:phases} are composed of periodic arrays of edge dislocations with distinct localized cores that contain different atomic densities in the GB. 
In the transition between these phases, dislocations of a less dense GB ($\mathbf{b}_{\text{I}}$, $n = 0$) pair up to form a new dislocation core ($\mathbf{b}_{\text{II}}$, $n = 0.5$), leading to a doubling of the Burgers vector and absorption of interstitial atoms. 
Previous studies of GB phase transitions in low-angle GBs revealed defect absorption by individual dislocation cores without the change of the Burgers vector~\cite{zhu_2018, frolov_2018_W} and other studies demonstrated the change in the dislocation network topology due to temperature~\cite{olmsted_2011} and solute segregation~\cite{sickafus_1987}.
It is also well established and expected that individual dislocations absorb point defects by climb~\cite{komem_1972};
yet here, we demonstrate a different mechanism where the dislocation network topology and the number of constituent atoms are coupled.
This coupling suggests an important mechanism for point-defect absorption in polycrystalline materials with non-equilibrium concentrations of point defects produced by rapid quenching, irradiation, or additive manufacturing approaches that can yield dense dislocation cellular walls~\cite{shang_2019}.
This work thus provides important insights into the ways in which low-angle GBs and dislocation arrays interact with point defects~\cite{beyerlein_2015}.
The work here, focused on an HCP metal, may be particularly relevant for engineering materials with such structures that experience radiation damage, such as Zr-based nuclear fuel cladding~\cite{yan_2015}.

Herein, we further extend the notion of the number of atoms in a GB plane ($N^{\mathrm{bulk}}_{\mathrm{plane}}$) to non-cubic, multi-basis crystals like HCP.
Previous studies on elemental cubic metals calculated this quantity as the total number of atoms located in one planar cut parallel to the GB, i.e., all these atoms are equidistant in the $z$-direction. 
This is not always the case for HCP metals or any multi-basis crystal, as visualized in \autoref{fig:terminations_Nplane}. 
Generally, $N^{\mathrm{bulk}}_{\mathrm{plane}}$ includes all atoms located inside a region with height equal to the minimum normal component of a lattice vector.
In this work, we show that if the GB structure search considers only those terminations of the surface by complete atomic layers, it is restricted to sampling states with $n=0$ or $n=0.5$ solely.
Importantly, we demonstrate that this restriction misses lower-energy GB structures with intermediate values of $n$.
A thorough search must consider all different atomic densities, as generalized by the framework presented here.

We implement this framework for handling the structural DOF and predicting new GB phases in the open-source GRIP tool, written in Python with minimal dependencies (see \nameref{sec:code_avail}).
The algorithm rapidly samples the configurational space described by relative translations and different atomic densities and moves the system toward equilibrium.
The relevant DOF---e.g., atomic density, reconstructions, temperature---are specified by the user in a single input file and the code exhaustively explores the GB phase space by sampling as many structures as possible in parallel.
The energy calculations presented here use empirical interatomic potentials (IAPs) to perform the dynamic sampling, but other techniques such as DFT can be used as well, as those calculations are decoupled from the structure optimization steps;
however, the use of IAPs enables us to access low-angle GBs and larger reconstructions with thousands of atoms, as demonstrated here in simulations up to $3\, \times\, 13$ reconstructions to validate the dislocation character.
This methodology can thus take advantage of the increasing availability of computational resources and the advent of high-fidelity machine-learned IAPs to enable quantum-accurate atomistic simulations of large systems with extended defects~\cite{freitas_2022}.
Advances in sampling and structure generation algorithms 
will further expand the diversity of results and the modular structure of the code enables different techniques to be easily plugged in.
Of particular interest would be extensions to multicomponent systems, which could be handled using a Monte Carlo approach~\cite{banadaki_2018, yang_2018} for compositional DOF and would enable grand canonical sampling of GB structures in technologically relevant alloy chemistries.

\section*{Methods}    \label{sec:methods}  

\subsection*{GB structure search}

\begin{figure}[!ht]
    \centering
    \includegraphics[width=0.5\linewidth]{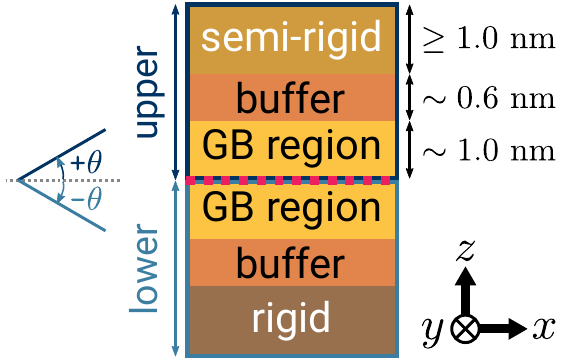}
    \caption{\textbf{Simulation cell setup for GRIP}.
    The cell is oriented such that the $y$-axis is the tilt axis direction, the $x$-axis is the orthogonal in-plane direction, and the $z$-axis is the out-of-plane normal direction.
    Periodic boundary conditions are maintained in the GB plane ($xy$-plane).
    GRIP begins by translating the upper crystal and removing atoms from the GB (magenta line).
    During MD, atoms in the GB regions are free to move while atoms in the buffer and semi-rigid regions in the upper slab are constrained to move together, and atoms in the lower two regions are fixed.
    During relaxation, atoms in both buffer regions are free to move while the semi-rigid region is still constrained.
    For the $\gamma$-surface method, the upper slab is only allowed to translate as a whole before relaxation is applied.
    }
    \label{fig:methods}
\end{figure}

We perform atomic-level optimization of GB structures using the open-source, Python-based {GRand} canonical Interface Predictor (GRIP) code (see \nameref{sec:code_avail}), which rigorously explores structural DOF through dynamic sampling.
Bicrystal slabs can be automatically generated using the Atomic Simulation Environment (ASE)~\cite{larsen_2017} library or supplied as external files, which we created for $\alpha$-Ti using a combination of ASE and Pymatgen~\cite{ong_2013}.
\autoref{fig:methods} shows the orientation of the simulation cell, such that the \mbox{$y$-axis} is the tilt axis direction, the \mbox{$x$-axis} is the orthogonal in-plane direction, and the \mbox{$z$-axis} is the out-of-plane normal direction.
We ensure periodicity in the GB plane (\mbox{$xy$-plane}) and an integer multiple of the interplanar spacing that totals at least \SI{3.5}{\nano\meter} in the $z$-direction for each slab to minimize cell size effects.
While any size cell can be used in principle, for computational tractability in this high-throughput study, we choose to simulate only \tz, \tx, and \ty\ STGBs where all Miller-Bravais indices for the plane and the in-plane $x$-direction are less than or equal to 15, resulting in 16, 40, and 94 STGBs for each of the tilt axes, respectively (150 total).

For an individual GB, each iteration of the algorithm has three stages.
During the first stage, the initial configuration is created by uniformly sampling a specific set of GB DOF. 
Specifically, the algorithm randomly samples an $m\, \times\, n$ replication of the unit GB cell (here, up to $3\, \times\, 3$), randomly translates the upper slab in the \mbox{$xy$-plane}, and removes a randomly chosen (from the user-specified interval) fraction of atoms from the GB. 
To further increase the structural diversity of the initial GB configurations, we have implemented random swaps of atoms on crystal lattice sites and interstitial sites in the GB region. 
The algorithm identifies interstitial sites near the GB as the vertices of the Voronoi diagram of the GB region.

During the second stage, it performs dynamic sampling to optimize the GB structure consistent with the imposed DOF. In this study, we used standard finite-temperature MD simulations using the Large-scale Atomic/Molecular Massively Parallel Simulator (LAMMPS)~\cite{thompson_2022} in the canonical ($NVT$) ensemble with a Langevin thermostat and a time step of \SI{2}{\femto\second} as our dynamic sampling technique.
The temperature and duration of the MD are also randomly sampled based on the user-specified ranges and only the atoms in the GB region are allowed to move freely during the dynamical sampling phase. It is straightforward to substitute this MD optimization with other, more sophisticated sampling techniques implemented in LAMMPS or other codes.
For this study, we choose a GB region of \SI{1}{\nano\meter} thickness on each side, a temperature between \SI{300}{\kelvin} and \SI{1200}{\kelvin}, and a duration up to \SI{0.6}{\nano\second}.
Finally, the temperature is quickly ramped down to \SI{100}{\kelvin} for \SI{2}{\pico\second}.
With some probability (here, \SI{5}{\percent}) the algorithm skips the dynamic sampling for one iteration and jumps to the third stage.

In the third stage, each GB structure following MD sampling is fully relaxed at \SI{0}{\kelvin} using a conjugate gradient minimization scheme, where atoms in the GB and buffer regions can move freely while the semi-rigid region is constrained to move together.
Here, we specified the buffer region to be \SI{0.6}{\nano\meter} beyond each side of the GB region, and larger values lowered $E_{\mathrm{gb}}$ by no more than \SI{1}{\percent}.
The convergence criteria are \SI{e-15} for relative energy ($\mathrm{d}E/E$ in successive iterations) and \SI{e-15}{\electronvolt\per\angstrom} for forces, with a maximum of \SI{e5} evaluations for each criterion.
The algorithm repeats these stages on each processor independently until termination, saving each relaxed structure to disk and periodically deleting duplicates.
Duplicates are defined as structures with the same value of $E_{\mathrm{gb}}$ and $n$ to three decimal places, and the algorithm will keep the structure with a smaller reconstruction and relative translations.

For each relaxed structure, the GB energy, $E_{\mathrm{gb}}$, is computed according to:

\begin{equation}
    E_{\mathrm{gb}} = \frac{E^{\mathrm{gb}}_{\mathrm{total}} - N^{\mathrm{gb}}_{\mathrm{total}} E^{\mathrm{bulk}}_{\mathrm{coh}}}{A^{\mathrm{gb}}_{\mathrm{plane}}}
    \label{eq:Egb}
\end{equation}

\noindent where $E^{\mathrm{gb}}_{\mathrm{total}}$ and $N^{\mathrm{gb}}_{\mathrm{total}}$ are the total energy and number, respectively, of atoms in the GB and buffer regions, $E^{\mathrm{bulk}}_{\mathrm{coh}}$ is the cohesive energy per atom in bulk $\alpha$-Ti, and $A^{\mathrm{gb}}_{\mathrm{plane}}$ is the area of the GB plane.
We also track the fraction of atoms in one plane or GB atomic density, $n$, according to:

\begin{equation}
    n = \frac{N_{\mathrm{total}} \mod N^{\mathrm{bulk}}_{\mathrm{plane}}}{N^{\mathrm{bulk}}_{\mathrm{plane}}} \in [0, 1)
    \label{eq:n}
\end{equation}

\noindent where $N_{\mathrm{total}}$ is the total number of atoms in the simulation cell and $N^{\mathrm{bulk}}_{\mathrm{plane}}$ is the number of atoms in one plane of the bulk structure.
Previous calculations of $N^{\mathrm{bulk}}_{\mathrm{plane}}$ simply counted the number of atoms at a single $z$ value in the bulk~\cite{frolov_2013_gbphase, zhu_2018};
however, due to the 2-atom basis of the HCP crystal structure, atoms associated with one plane may be offset in the $z$-direction, as we show in \autoref{fig:terminations_Nplane}.
Therefore, we calculate $N^{\mathrm{bulk}}_{\mathrm{plane}}$ as the number of atoms within a region equal to the minimum normal component of a lattice vector; in HCP $\alpha$-Ti, this is equivalent to the interplanar spacing of the hexagonal lattice ($d_{hkl}$) given by~\cite{cullity_2001}:

\begin{equation}
    \frac{1}{d_{hkl}^2} = \frac{4}{3} \left( \frac{h^2 + hk + k^2}{a^2} \right) + \left( \frac{l}{c} \right)^2
    \label{eq:dhkl}
\end{equation}

\noindent where $h$, $k$, and $l$ are the Miller indices, and $a$ and $c$ are the HCP lattice constants.
We note this extended definition of $N^{\mathrm{bulk}}_{\mathrm{plane}}$ reduces to taking a planar slice for unary, single-basis systems like elemental BCC and FCC metals, consistent with previous studies~\cite{frolov_2013_gbphase, zhu_2018}.

We compare the results of structure optimization using two different interatomic potentials, an embedded-atom method (EAM) potential for Ti--Al from Zope and Mishin~\cite{zope_2003} and a modified embedded-atom method (MEAM) potential for Ti from Hennig, et al.~\cite{hennig_2008}
For each STGB and potential, we also optimize the structure using the $\gamma$-surface method~\cite{mishin_1998} for comparison, using a $2\, \times\, 4$ replication of the same bicrystals and translating the top slab in increments of \SI{0.025}{\nano\meter} in the $x$- and $y$-directions prior to a conjugate gradient energy minimization.

\subsection*{High-temperature MD simulations}

To study GB phase stability and transitions, we perform high-temperature MD simulations using methods adapted from previous work~\cite{frolov_2013_gbphase}.
Briefly, we replicate the STGB in the $x$- and $y$-directions until the simulation cell is around \SI{10}{\nano\meter} in the $x$-direction and \SI{3}{\nano\meter} in the $y$-direction along the tilt axis.
We freeze the bottom \SI{1}{\nano\meter} layer of atoms and constrain the top \SI{1}{\nano\meter} layer to be semi-rigid throughout the \SI{20}{\nano\second} simulation.
We use periodic boundary conditions (PBCs) in the $y$-direction and both PBCs and open surfaces with \SI{1}{\nano\meter} of vacuum in the $x$-direction.
We scan a range of temperatures between \SI{600}{\kelvin} and \SI{1200}{\kelvin}.

To induce a phase transition, we either insert additional Ti atoms at interstitial sites in the GB region or delete Ti atoms from a region near the top of the GB region.
These MD simulations are performed in the canonical ($NVT$) ensemble between \SI{600}{\kelvin} and \SI{1200}{\kelvin} for up to \SI{20}{\nano\second}, using the MEAM potential and associated structures.
For clarity of visualization, we relax all structures at \SI{0}{\kelvin} using a conjugate gradient minimization scheme.

\subsection*{DFT calculations}

To validate select GB structures, we perform additional density functional theory (DFT) calculations using the Vienna Ab initio Simulation Package (VASP)~\cite{kresse_1993, kresse_1994, kresse_1996a, kresse_1996b} with projector augmented-wave potentials~\cite{kresse_1999} and the generalized gradient approximation exchange correlation functional of Perdew, Burke, and Ernzerhof~\cite{perdew_1996}.
The semi-core $3p$ states are treated as valence states (\texttt{Ti\_pv} potential).
We use Monkhorst-Pack~\cite{monkhorst_1976} $\mathbf{k}$-point grids with a density of 5000 $\mathbf{k}$ points per reciprocal atom and apply Methfessel-Paxton smearing~\cite{methfessel_1989} with a width of \SI{0.1}{\electronvolt}.
The plane wave cutoff energy is \SI{500}{\electronvolt} and the convergence criteria are set at \SI{e-5}{\electronvolt} for energy and \SI{0.02}{\electronvolt\per\angstrom} for forces.
We create the input structure by extracting a section near the GB region of the optimized structure from GRIP of approximately \SI{4.5}{\nano\meter} in thickness (200--300 atoms) and adding \SI{1}{\nano\meter} of vacuum on top. 
The axes are rescaled to match equilibrium DFT values and atomic positions are fully relaxed while the cell shape and volume are fixed to maintain stresses in the GB plane.
The energy of the GB is computed as the difference in total energy of a structure with the GB and a second bulk structure in the same orientation with the same number of atoms and vacuum but without a GB, divided by the planar area.

\section*{Data availability}   \label{sec:data_avail}
The data that support the findings of this study, including input and relaxed STGB structures, are available at Zenodo at publication time.
Other data are available from the corresponding author upon reasonable request.

\section*{Code availability}   \label{sec:code_avail}
The {GRand} canonical Interface Predictor (GRIP) tool that implements the GB structure optimization algorithm described here can be found at \url{https://github.com/enze-chen/grip} at publication time.

\section*{Acknowledgments}
This work was performed under the auspices of the U.S. Department of Energy (DOE) by the Lawrence Livermore National Laboratory (LLNL) under Contract No. DE-AC52-07NA27344. 
Computing support for this work comes from the LLNL Institutional Computing Grand Challenge program and Bridges-2 at Pittsburgh Supercomputing Center through allocation DMR110087 from the Advanced Cyberinfrastructure Coordination Ecosystem: Services \& Support (ACCESS) program.
T.F. acknowledges support from the U.S. DOE, Office of Science under an Office of Fusion Energy Sciences Early Career Award.
E.C. and M.A. acknowledge funding through the DOE, Office of Science, Office of Basic Energy Sciences, Materials Sciences and Engineering Division, under Contract No. DE-AC02-05-CH11231 within the Materials Project program (KC23MP).
E.C. also acknowledges a fellowship through the National Science Foundation Graduate Research Fellowship Program under Grant No. DGE-2146752 and support from the Computational Chemistry \& Materials Science Institute at LLNL.
Part of this work was funded by the Laboratory Directed Research and Development (LDRD) program at LLNL under projects with a tracking code 22-ERD-002. 
Part of this work was also supported by the U.S. DOE, Office of Energy Efficiency and Renewable Energy, Hydrogen and Fuel Cell Technologies Office, through the Hydrogen storage Materials Advanced Research Consortium (HyMARC).
The authors thank Tomas Oppelstrup and Daryl Chrzan for helpful discussions.
All figures are produced using matplotlib\cite{hunter_2007} and OVITO~\cite{stukowski_2009}.

\section*{Author contributions}
T.F. and E.C. designed the study which was initially supervised by T.W.H. and B.C.W.
T.F. and E.C. designed the GCO algorithm.
E.C. performed the simulations and analyzed the results.
E.C. and T.F. drafted the manuscript with feedback from M.A.
All authors discussed the results and contributed to the writing of the manuscript.

\section*{Declaration of competing interests}
The authors declare no competing financial or non-financial interests.

\section*{Materials and correspondence}
Correspondence and requests for materials should be addressed to E.C. or T.F.

\nolinenumbers
\singlespacing
\small 
\bibliography{main}


\newpage 
\normalsize 
\linenumbers

\begin{center}
	{\large 
		\textbf{Supplementary Information for \\ Grand canonically optimized grain boundary phases in hexagonal close-packed titanium}
		\vspace{0.2in}}
	
	{Enze Chen$^{1,2,3,4,*}$, Tae Wook Heo$^{2}$, Brandon C. Wood$^{2}$, Mark Asta$^{1,3}$, Timofey Frolov$^{2, *}$}
\end{center}

{\noindent \fontsize{10}{12}\selectfont 
	$^1$ Department of Materials Science and Engineering, University of California, Berkeley, CA 94720, USA \\
	$^2$ Materials Science Division, Lawrence Livermore National Laboratory, Livermore, CA 94550, USA \\
	$^3$ Materials Sciences Division, Lawrence Berkeley National Laboratory, Berkeley, CA 94720, USA \\
	$^4$ Present address: Department of Materials Science and Engineering, Stanford University, Stanford, CA 94305, USA \\
	Email: 
	$^{*}$enze@stanford.edu; 
	$^{*}$frolov2@llnl.gov
}

\setcounter{figure}{0}
\renewcommand{\thefigure}{S\arabic{figure}}
\setcounter{table}{0}
\renewcommand{\thetable}{S\arabic{table}}
\setcounter{equation}{0}
\renewcommand{\theequation}{S\arabic{equation}}
\vspace{20pt}

\begin{figure}[!ht]
	\centering
	\includegraphics[width=\linewidth]{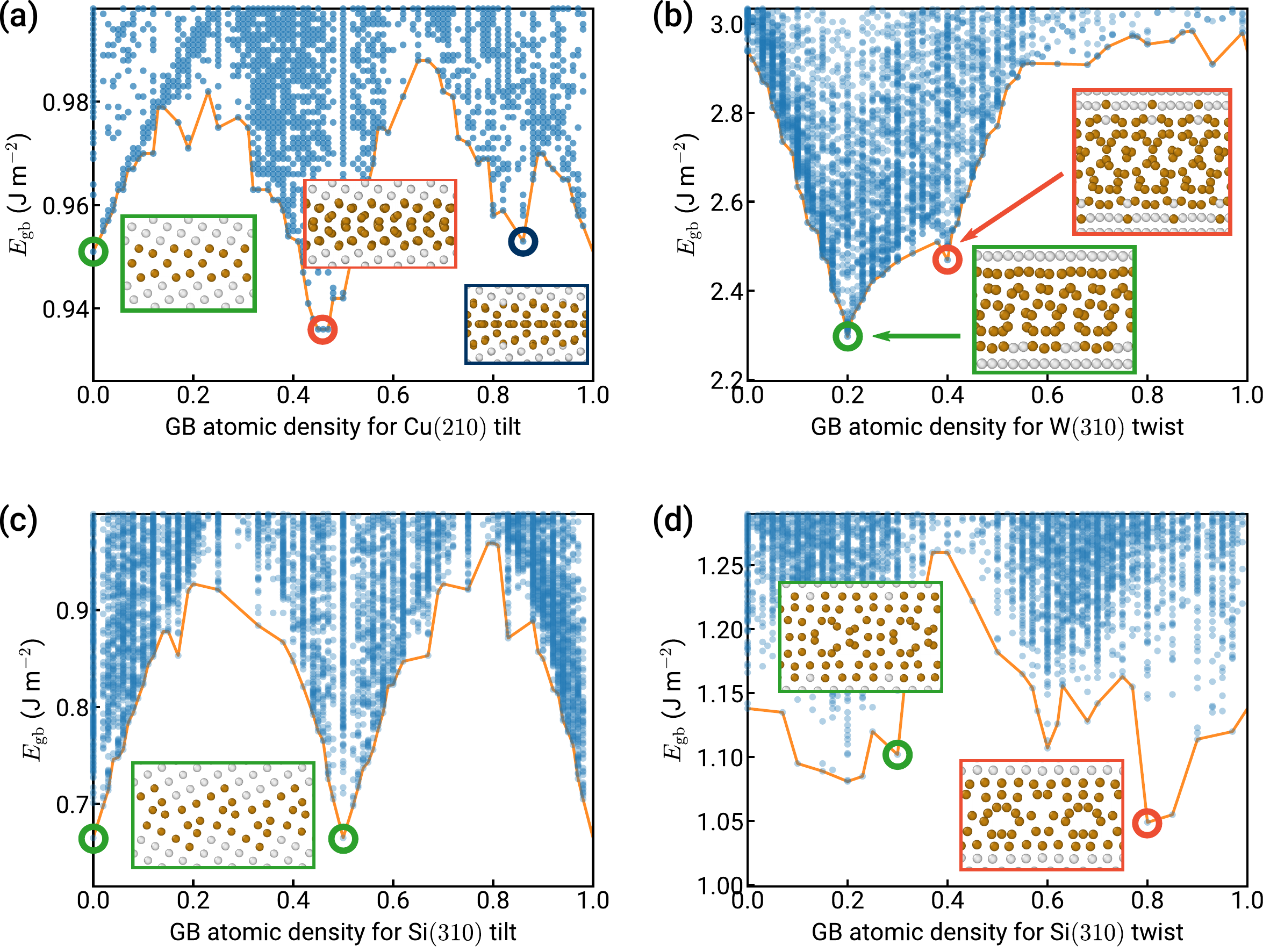}
	\caption{\textbf{Validation of the GRIP tool}.
		We perform grand canonical optimization of the 
		(a) $\Sigma 5 (210) [001]$ tilt GB in Cu (EAM potential~\cite{mishin_2001}), 
		(b) $\Sigma 5 (310) [001]$ twist GB in W (EAM~\cite{zhou_2004}),
		(c) $\Sigma 5 (310) [001]$ tilt and (d) twist GB in Si (Stillinger-Weber~\cite{stillinger_1985}).
		The $E_{\mathrm{gb}}$ vs. $n$ plots and low-energy structures match those in the literature for Cu~\cite{frolov_2013_gbphase}, W~\cite{frolov_2018_W}, and Si~\cite{banadaki_2018, vonalfthan_2006}.
		The atoms are colored according to CNA~\cite{stukowski_2012_cna}, where gray are bulk-coordinated atoms and brown are non-bulk-coordinated atoms (in the GB).
	}
	\label{fig:gcoval1}
\end{figure}

\newpage 
\begin{figure}[!ht]
	\centering
	\includegraphics[width=\linewidth]{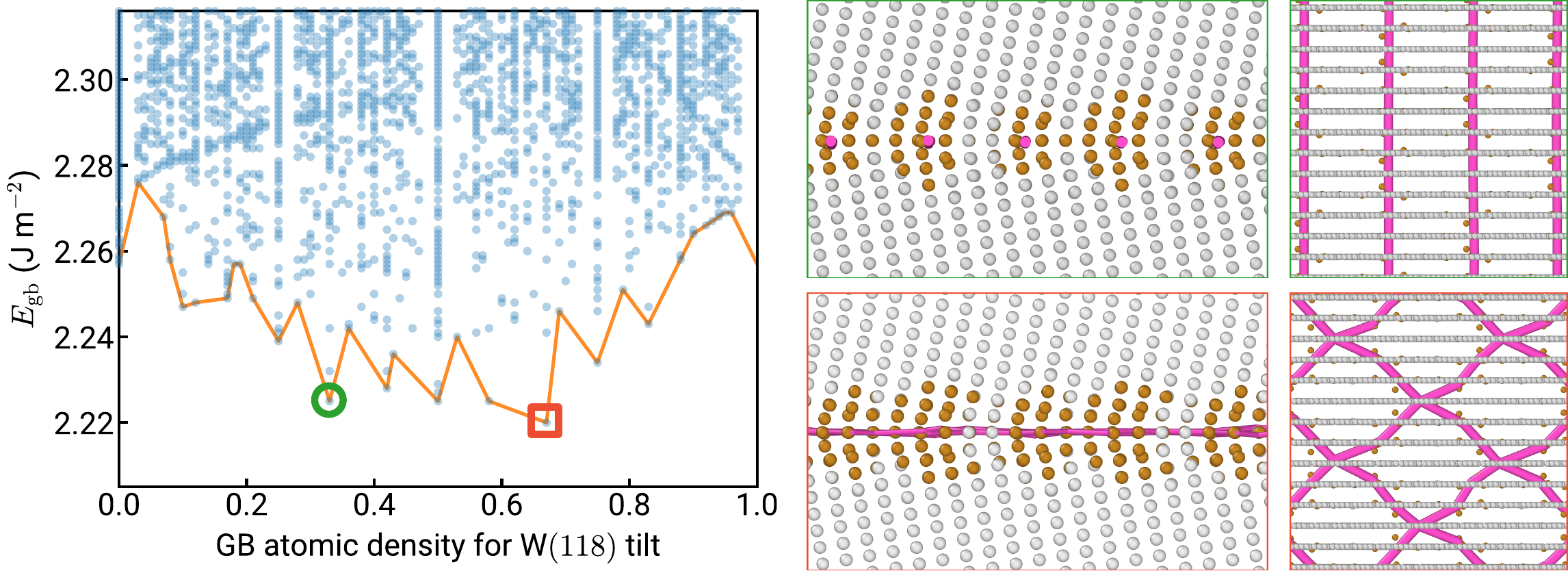}
	\caption{\textbf{Discovering a new ground state in W$(118)[1\bar{1}0]$}.
		A previous study~\cite{frolov_2018_W} using the evolutionary algorithm USPEX~\cite{oganov_2006, lyakhov_2013} and an EAM potential~\cite{zhou_2004} found a ground state for the W$(118)[1\bar{1}0]$ STGB at $n = 0.33$ with $E_{\mathrm{gb}} = \SI{0.225}{\joule\per\meter\squared}$, as pictured on the top.
		Using GRIP and the same EAM potential, we find a structure with lower energy $E_{\mathrm{gb}} = \SI{0.220}{\joule\per\meter\squared}$ at $n = 0.67$, as shown on the bottom.
		This structure has a different dislocation network where the $\langle 001 \rangle$-type edge dislocations (magenta, as identified using DXA) overlap in the GB plane instead of residing in parallel.
	}
	\label{fig:gcoval2}
\end{figure}

\newpage 
\begin{figure}[!ht]
	\centering
	\includegraphics[width=\linewidth]{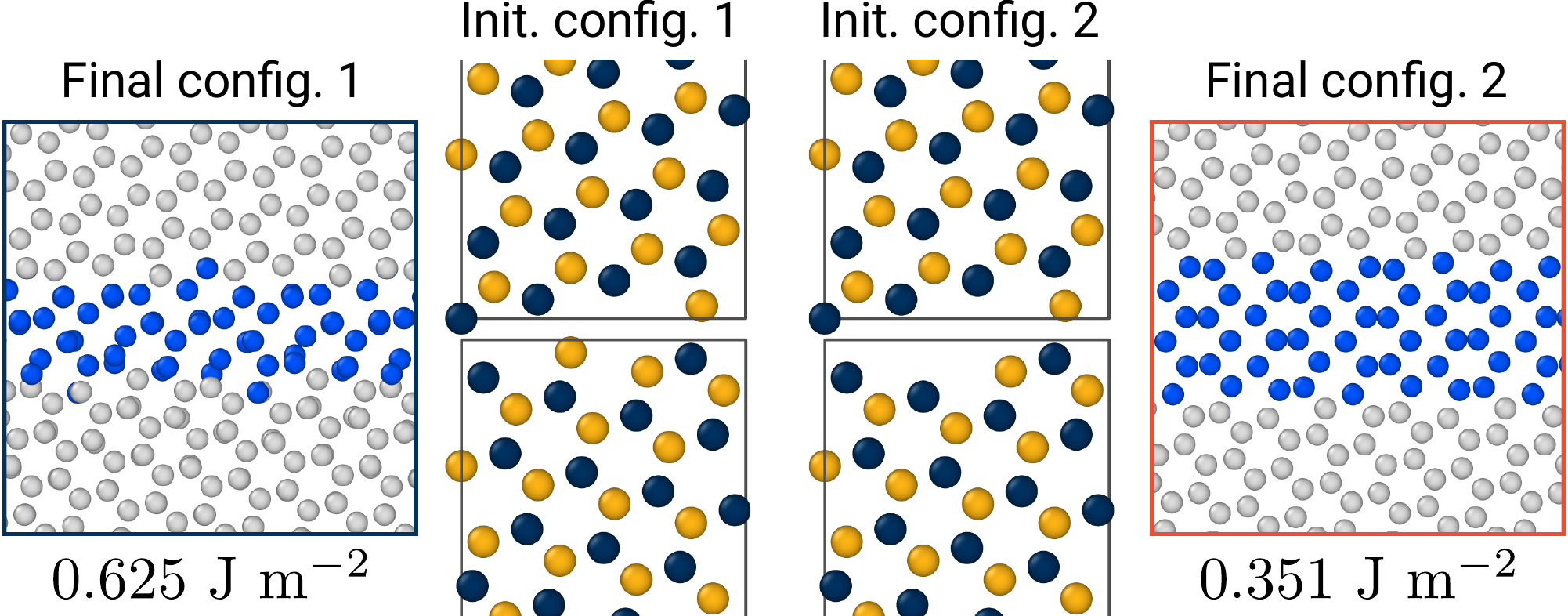}
	\caption{\textbf{Different planar terminations and corresponding GB structures}.
		$\{ 3 1 \bar{4} 0 \}$\tz\ is a GB where the two basis atoms have different $z$ coordinates normal to the GB plane, and two initial configurations are shown where the lower slabs terminate at different basis atoms (gold for the first, blue for the second).
		We use the $\gamma$-surface method to optimize the GB structure in both cases.
		The first configuration ($n = 0$) produces a higher-energy structure shown on the left ($E_{\mathrm{gb}} = \SI{0.625}{\joule\per\meter\squared}$), while the second configuration ($n = 0.5$) produces the lower-energy structure shown on the right ($E_{\mathrm{gb}} = \SI{0.351}{\joule\per\meter\squared}$), which matches the ground-state structure from GRIP.
		The final structures and colors correspond to those in Figure 3 in the main manuscript.}
	\label{fig:terminations}
\end{figure}

\newpage

\begin{figure}[!ht]
	\centering
	\includegraphics[width=\linewidth]{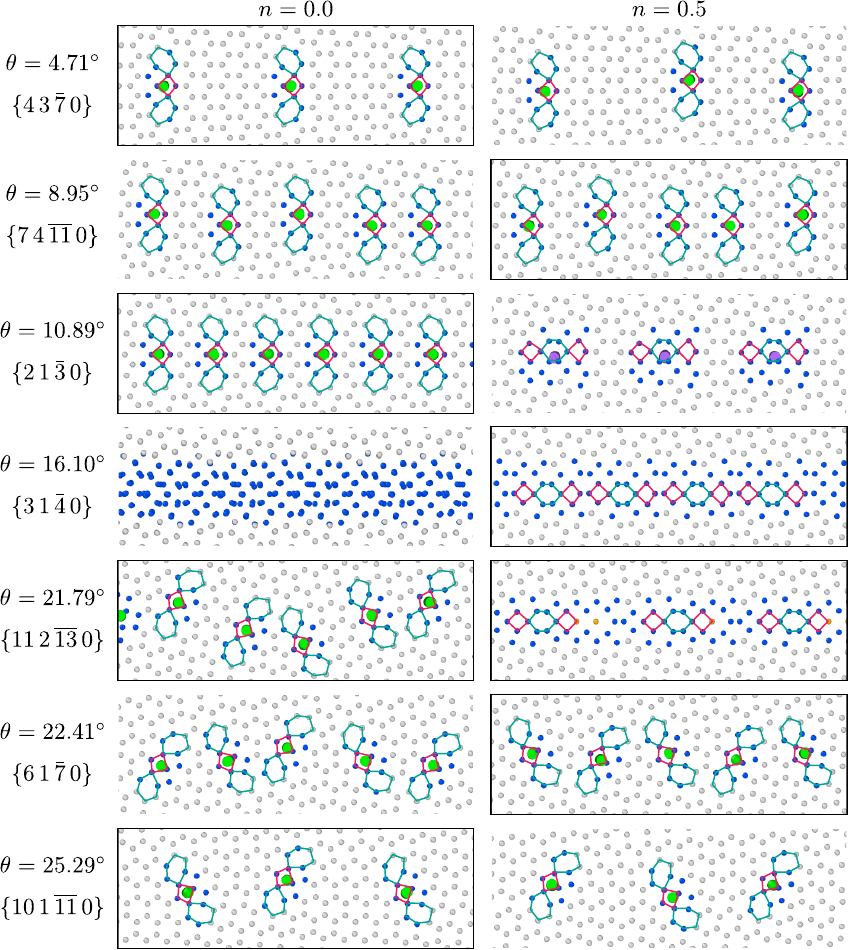}
	\caption{\textbf{GB structural units in \tz\ STGBs}.
		GRIP results for select GBs at $n=0.0$ and $n=0.5$ viewed down \tz.  
		Low-angle GBs adopt a dislocation core configuration at $n = 0.0$ that is accommodated at $n=0.5$ through dislocation climb.
		At $\theta \approx \SI{10.89}{\degree}$, there is a transition at $n=0.5$ to a different structural unit.
		At even higher angles $\theta \ge \SI{22.41}{\degree}$, the GB structural units transform back into the motifs at low angles.
		The lower-energy structure at each tilt angle, as evaluated using the MEAM potential~\cite{hennig_2008}, is outlined in black.
	}
	\label{fig:comp_units}
\end{figure}

\newpage
\begin{figure}[!ht]
	\centering
	\includegraphics[width=0.6\linewidth]{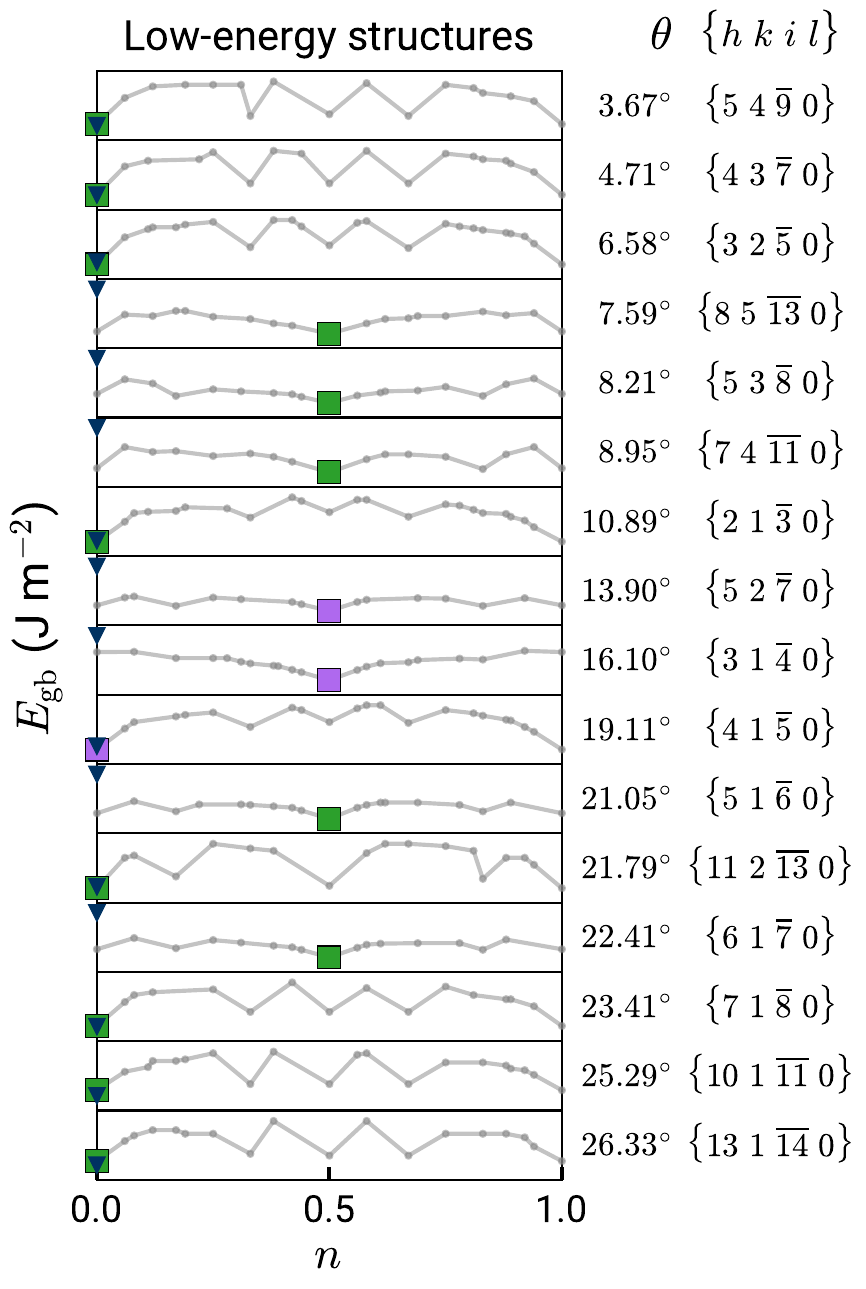}
	\caption{\textbf{Energy map of \tz\ STGBs simulated using an EAM potential}~\cite{zope_2003}.
		The profiles are qualitatively similar to those in Figure 4 in the main text, which was generated using a MEAM potential~\cite{hennig_2008}.
		See Figure 4 for additional descriptions of features.
	}
	\label{fig:Eminima_Zope_1}
\end{figure}

\newpage 
\begin{figure}[!ht]
	\centering
	\includegraphics[width=\linewidth]{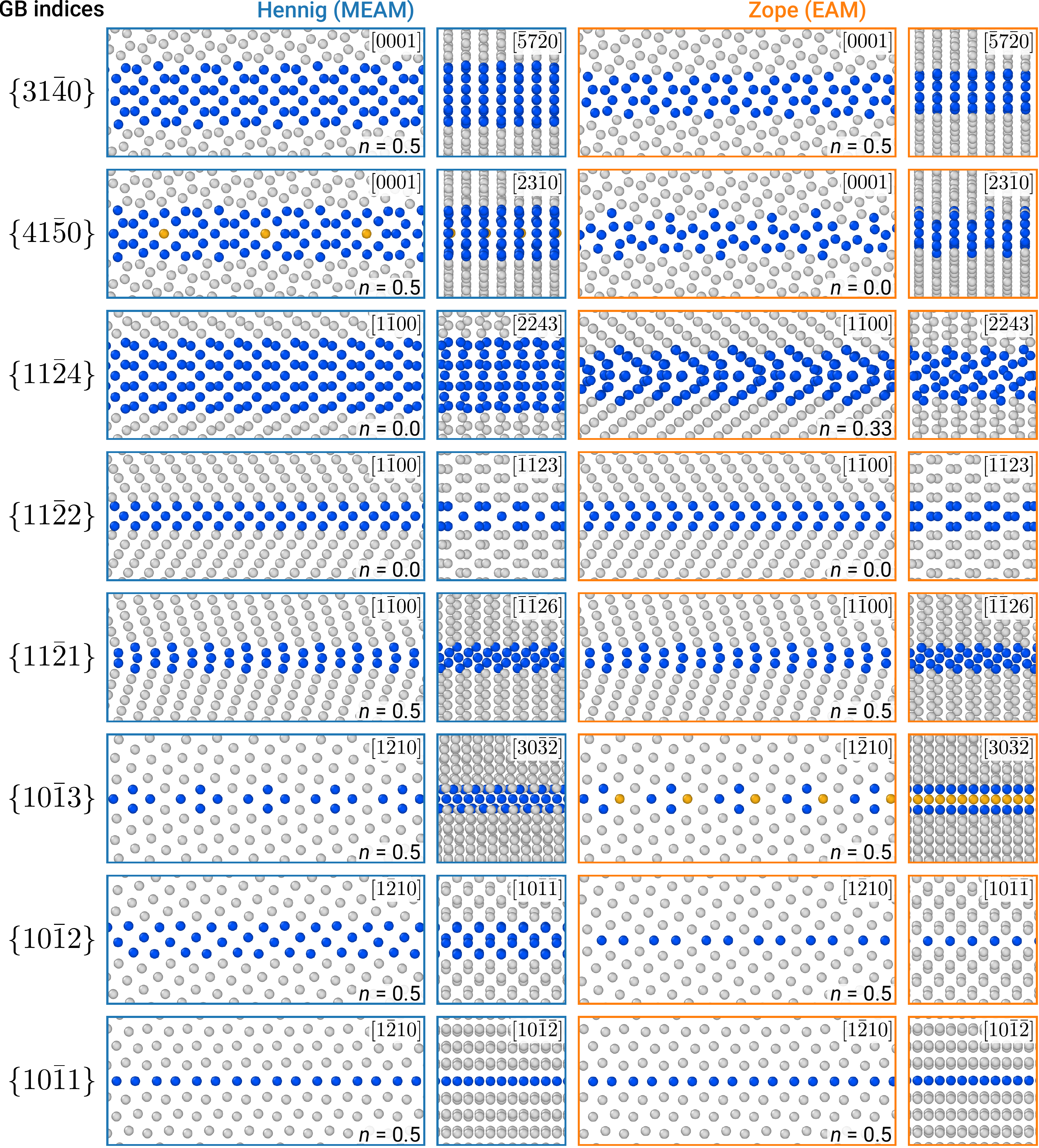}
	\caption{\textbf{Comparison of optimized GB structures}.
		A few STGBs are provided for each tilt axis and two projections are shown for each ground state obtained using the EAM~\cite{zope_2003} and MEAM~\cite{hennig_2008} potentials.
		The GB atomic density ($n$) is shown in the lower-right corner and may not be equal for both potentials.
		The atoms are colored according to the common neighbor analysis (CNA) in OVITO~\cite{stukowski_2012_cna, stukowski_2009}, where gray are HCP-coordinated atoms, gold are FCC-coordinated atoms, and blue have a different coordination (in the GB).
	}
	\label{fig:comp_structs}
\end{figure}

\newpage 
\begin{figure}[!ht]
	\centering
	\includegraphics[width=0.8\linewidth]{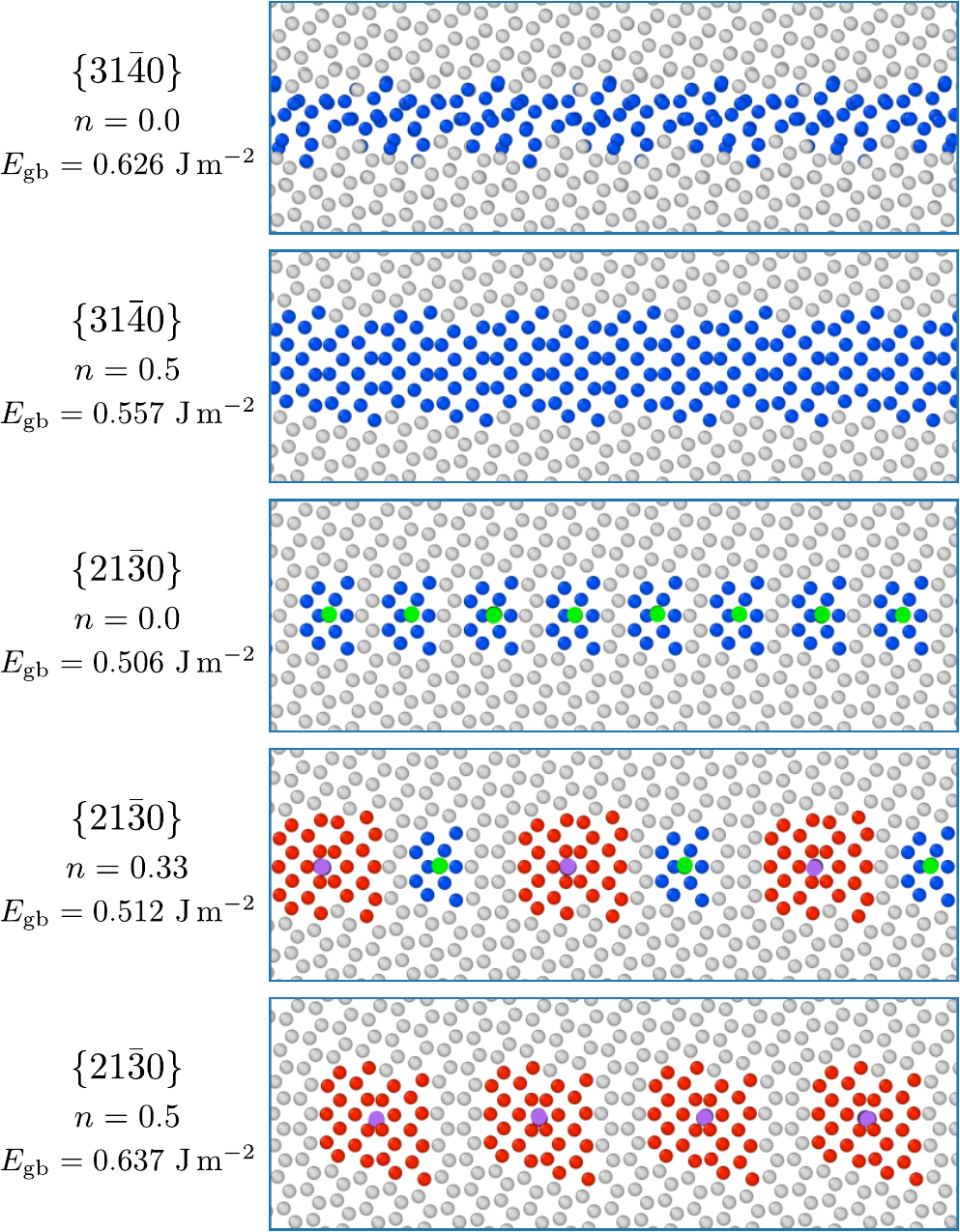}
	\caption{\textbf{DFT validation for select STGBs}.
		The structures from GRIP are used as inputs to VASP for further relaxation (see Methods in the main manuscript for details).
		The GB dislocation core structures remain stable and the relative energies between different phases are consistent with those from GRIP.
	}
	\label{fig:dft_val}
\end{figure}

\newpage
\begin{figure}[!ht]
	\centering
	\includegraphics[width=\linewidth]{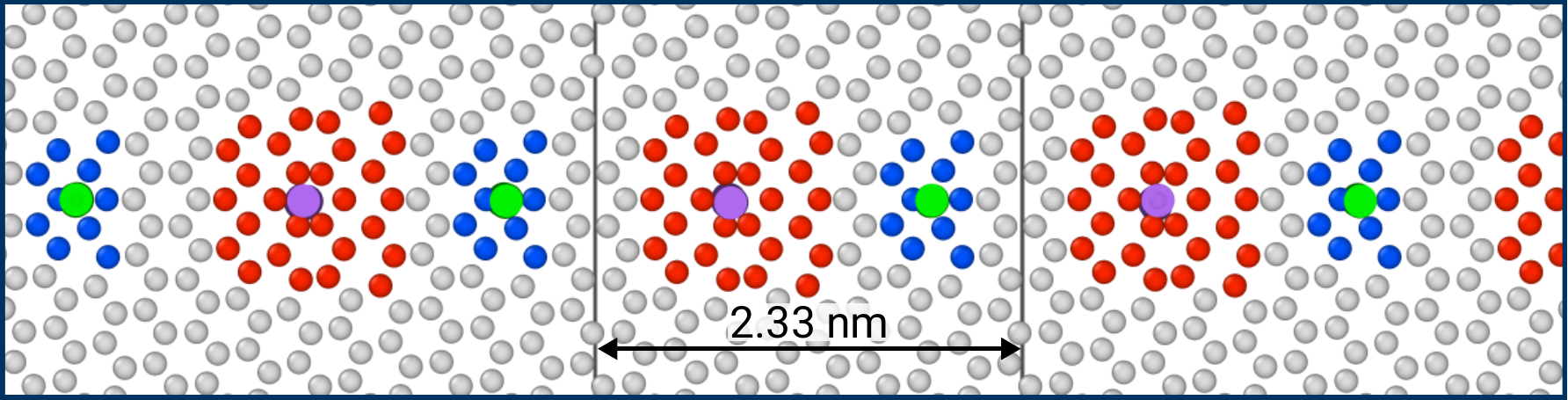}
	\caption{\textbf{Mixed state at $n = 0.33$ and $n = 0.67$ for \gbb}.
		At intermediate values of $n \in (0, 0.5)$, the localized dislocation cores in the GB alternate between $\mathbf{b}_{\text{I}} = \frac{1}{3} \langle 1 \bar{2} 1 0 \rangle$ (green) and $\mathbf{b}_{\text{II}} = \frac{1}{3} \langle 2 \bar{4} 2 0 \rangle$ (purple) dislocations to obtain the minimum-energy structure.
		The amount of each phase follows the conventional lever rule for phase fractions.
		Dislocations are identified using the dislocation extraction algorithm (DXA) in OVITO~\cite{stukowski_2012_dxa, stukowski_2009}.
	}
	\label{fig:phases_mixed}
\end{figure}

\vspace{10ex}

\begin{figure}[!ht]
	\centering
	\includegraphics[width=\linewidth]{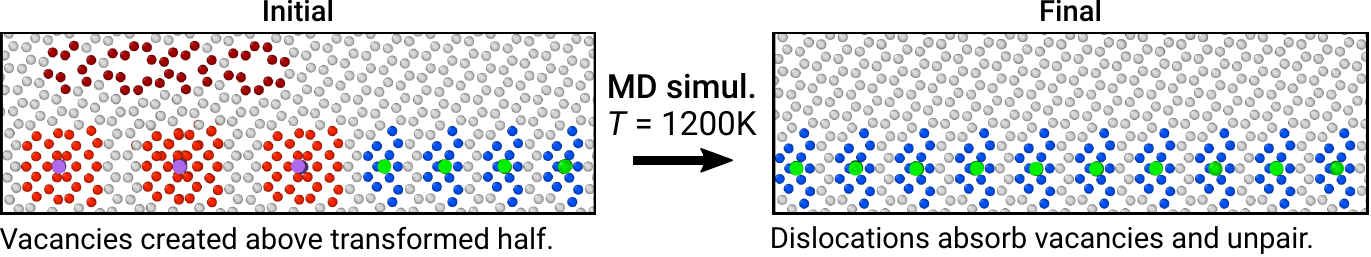}
	\caption{\textbf{(Reverse) phase transformation through vacancy absorption}.
		In the main text, we demonstrate interstitial-induced phase transformation and coexistence in \gbb\tz, where every two $\mathbf{b}_{\text{I}}$ dislocations pair up to form one $\mathbf{b}_{\text{II}}$ dislocation.
		Here, by injecting vacancies (outlined in dark red) and performing MD simulations at \SI{1200}{\kelvin}, we reverse the transformation, whereby the $\mathbf{b}_{\text{II}}$ dislocations absorb the vacancies and unpair.
	}
	\label{fig:phases_si}
\end{figure}


\newpage
\doublespacing

\section*{Supplementary Note 1}   \label{sec:si_comp}

\begin{figure}[!ht]
	\centering
	\includegraphics[width=\linewidth]{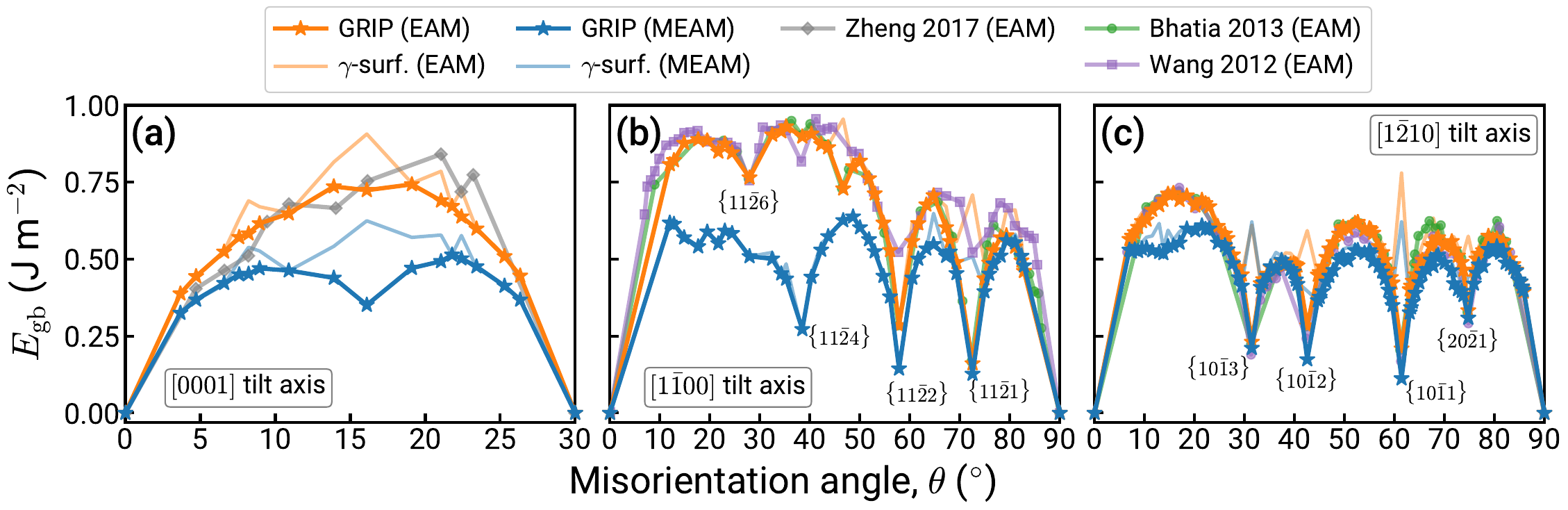}
	\caption{\textbf{$E_{\mathrm{gb}}$ vs. $\theta$ for STGBs in $\alpha$-Ti}.
		All results are shown for the (a) \tz, (b) \tx, and (c) \ty\ tilt axes.
		Each point corresponds to the minimum-energy structure for that tilt angle.
		The orange and blue lines correspond to our calculations using the EAM~\cite{zope_2003} and MEAM~\cite{hennig_2008} potentials, respectively.
		Solid lines with stars are for GRIP and translucent lines are for the $\gamma$-surface method.
		The other data (gray in (a) and green/purple in (b) and (c)) are referenced from the literature~\cite{zheng_2017, bhatia_2013, wang_2012a, wang_2012b}.
		Select TBs corresponding to energy cusps are labeled in panels (b) and (c).
	}
	\label{fig:misorientation}
\end{figure}

To better characterize where our algorithm improves upon existing studies, we compare plots of $E_{\mathrm{gb}}$ vs. the tilt angle ($\theta$) for all ground-state structures in \autoref{fig:misorientation}.
Using the EAM potential from Zope and Mishin~\cite{zope_2003} (solid orange), we match or improve upon the results from Bhatia and Solanki~\cite{bhatia_2013} (green) and Wang and Beyerlein~\cite{wang_2012a, wang_2012b} (purple) for the \tx\ and \ty\ STGBs, as seen in panels (b) and (c).
Our results for \tz\ STGBs are in good agreement with those from Zheng, et al.~\cite{zheng_2017} (gray), although more precise comparisons are not possible as they used a different EAM potential~\cite{ackland_1992}.
As noted in the main manuscript, previous studies may have inconsistently sampled different terminations~\cite{ni_2015} or deleted overlapping atoms~\cite{bhatia_2013}, so we also perform standard $\gamma$-surface calculations with perfect bulk slabs for each STGB and plot the results in corresponding translucent colors in \autoref{fig:misorientation}.
For both the EAM (orange) and MEAM (blue) potentials, the GRIP data (solid lines) are lower bounds for the $\gamma$-surface values (translucent), which is consistent with expectations.
We attribute the large differences in $\gamma$-surface sampling, i.e., the sharp peaks in panels (b) and (c), to our definition of planar terminations that may not have been similarly enforced in previous works~\cite{wang_2012a, wang_2012b, bhatia_2013, ni_2015}.
The majority of remaining discrepancies between the GRIP and $\gamma$-surface data occur for the family of \tz\ STGBs, with the largest difference at $\theta \approx \SI{16.1}{\degree}$, which is the \gba\ STGB shown in Figure 3 in the main manuscript.
We find for the other two tilt axes closer agreement for $E_{\mathrm{gb}}$, even when many STGBs require GCO, as seen in Figure 4 in the main text, suggesting that structural differences in these STGBs are small (see \autoref{fig:comp_structs}).
Consistent with the only existing study that used an evolutionary algorithm to study a few GBs in an HCP metal (Mg)~\cite{yang_2020}, we also observe a zigzag distribution of GB dislocations instead of a straight line in several \tx\ and \ty\ STGBs.

\begin{figure}[!hb]
	\centering
	\includegraphics[width=\linewidth]{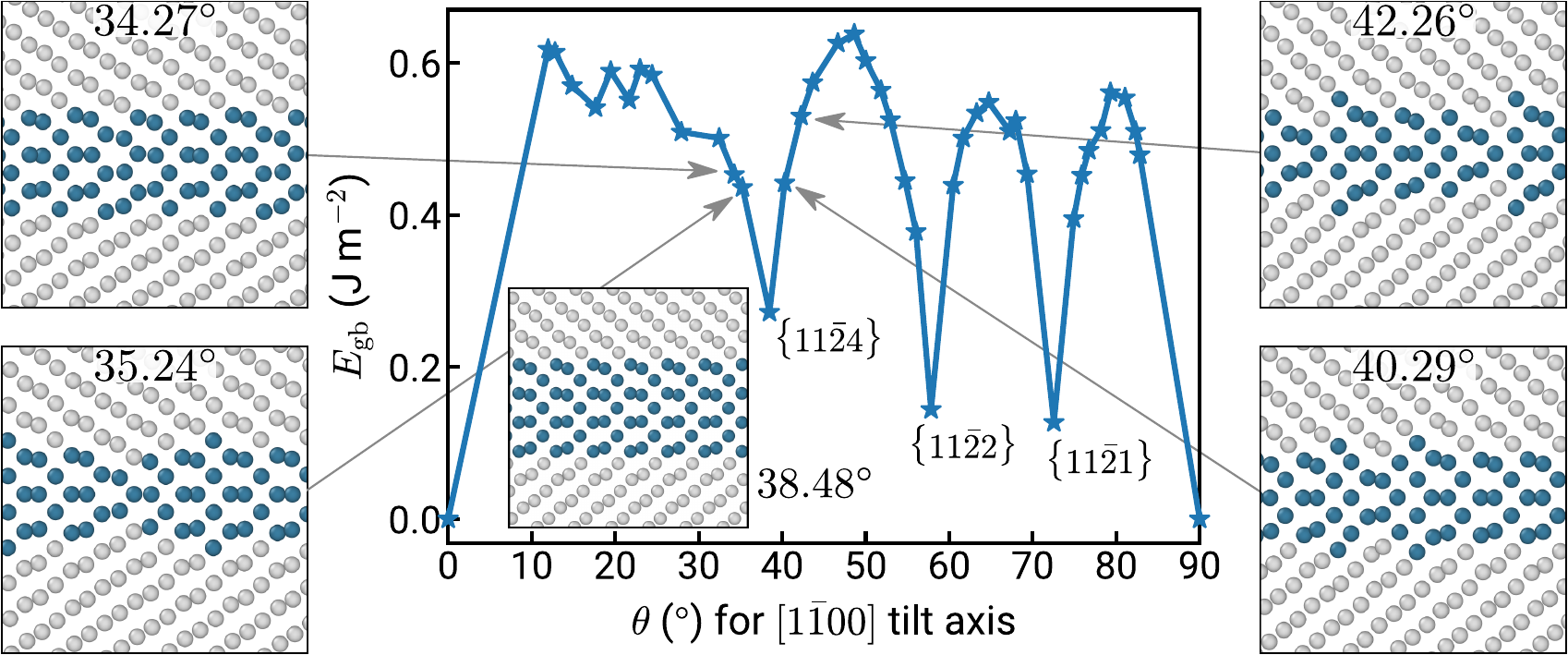}
	\caption{\textbf{Faceting of the boundary in \tx\ STGBs}.
		The $\{ 1 1 \bar{2} 4 \}$\tx\ ($\theta \approx \SI{38.5}{\degree}$) TB when simulated with the MEAM potential adopts a thick interfacial structure that is a strained version of a metastable bulk polymorph~\cite{hooshmand_2021, zarkevich_2016}.
		At nearby misorientation angles, the interfacial phase is partly preserved and the boundaries with the surrounding $\alpha$-Ti slabs are faceted.
	}
	\label{fig:faceting}
\end{figure}

Different empirical potential formalisms are expected to result in different GB properties, but previous benchmark studies on STGBs in cubic metals~\cite{waters_2023} and HCP $\alpha$-Zr~\cite{torres_2021} found largely similar $E_{\mathrm{gb}}$ vs. $\theta$ profiles using the $\gamma$-surface method.
In contrast, we find notable differences for multiple tilt axes in $\alpha$-Ti when comparing the EAM~\cite{zope_2003} (orange) and MEAM~\cite{hennig_2008} (blue) parameterizations.
In the family of \tz\ STGBs, the values for $E_{\mathrm{gb}}$ from the MEAM potential are consistently lower than those produced by the EAM potential, with a noticeable energy cusp (local minimum) at $\theta \approx \SI{16.1}{\degree}$.
Moreover, this cusp is only present for the GCO data and not the $\gamma$-surface results---which in fact peaks---further demonstrating the advantages of the GRIP algorithm.
Likewise, in the family of \tx\ STGBs (\autoref{fig:misorientation}b), only using the MEAM potential do we recover a low-energy $\{ 1 1 \bar{2} 4 \}$ TB at $\theta \approx \SI{38.5}{\degree}$ that we have extensively characterized using transmission electron microscopy and DFT~\cite{hooshmand_2021}.
There, we found the TB to adopt a thick body-centered orthorhombic (BCO) structure that is a strained version of a metastable bulk polymorph of Ti~\cite{zarkevich_2016};
however, what this high-throughput study also reveals is faceting around the interfacial phase at nearby tilt angles to accommodate the strained BCO phase, as shown in \autoref{fig:faceting}.
Stabilization of the BCO phase may be responsible for the significantly lower energy of \tx\ STGBs simulated with the MEAM potential vs. the EAM potential shown in \autoref{fig:misorientation}b, and we leave this analysis to future work.

\end{document}